\documentclass[useAMS,usenatbib,usegraphicx]{mn2e}

\usepackage{aas_macros}

\usepackage{color}

\usepackage{lineno}

\def\gx{GX~339$-$4}

\def\cx{Cyg~X$-$1}

\def\grs{GRS~1915$+$105}

\def\gro{GRO~J1655$-$40}

\def\1e{1E~1740.7$-$2942}

\def\xte{XTE~J1550$-$564}

\def\xt{XTE~J1650$-$500}

\def\vq{V404~Cyg}

\def\a0{A~0620$-$00}

\def\1h{H~1743$-$322}

\def\4u{4U~1755$-$33}

\def\sw{Swift J1753.5$-$0127}

\def\igr{IGR J17497$-$2821}

\def\maxisz{MAXI~J1659$-$152}

\def\xteds{XTE~J1752$-$223}

\def\ledd{\hbox{L$_{\rm Edd}$}}

\def\ergs{\hbox{erg s$^{-1}$}}

\def\ergsc{\hbox{erg s$^{-1}$ cm$^{-2}$}}

\title[Long term radio/X-ray flux correlation in \gx ]{The  ``universal" radio/X-ray flux correlation : the case study of the black hole \gx.  }
\author[S. Corbel et al.]{S. Corbel$^{1,2}$\thanks{E-mail:stephane.corbel@cea.fr} , 
M. Coriat$^{3,1}$, C. Brocksopp$^{4}$, A.K. Tzioumis$^{5}$,  R. P. Fender$^{3}$,
\newauthor 
J.A. Tomsick$^{6}$,   M. M. Buxton$^{7}$, C. D.  Bailyn$^{7}$ \\
$^{1}$Laboratoire AIM (CEA/IRFU - CNRS/INSU - Universit\'e Paris Diderot), CEA DSM/IRFU/SAp, F-91191 Gif-sur-Yvette, France\\
$^{2}$Institut Universitaire de France, 75005 Paris, France.  \\
$^{3}$School of Physics and Astronomy, University of Southampton, Highfield, Southampton, SO17 1BJ, UK. \\
$^{4}$Mullard Space Science Laboratory, University College London, Holmbury St Mary, Dorking, Surrey RH5 6NT, UK.\\
$^{5}$Australia Telescope National Facility, CSIRO, P.O. Box 76, Epping NSW 1710, Australia.\\
$^{6}$Space Sciences Laboratory, 7 Gauss Way, University of California, Berkeley, CA 94720-7450, USA.\\
$^{7}$Astronomy Department, Yale University, P.O. Box 208101, New Haven, CT 06520-8101, USA.
}
\begin{document}


\pagerange{\pageref{firstpage}--\pageref{lastpage}} \pubyear{2008}

\maketitle

\label{firstpage}

\begin{abstract}

The existing radio and X-ray flux correlation for Galactic black holes in the hard and quiescent states relies on a sample which is
mostly dominated by two sources (\gx\ and \vq ) observed in a single outburst.   
In this paper, we report on a series of  radio and X-ray  observations of the recurrent black hole \gx\ with the Australia Telescope Compact  Array,  the Rossi 
X-ray Timing Explorer and the Swift satellites. With our new  long term campaign, we now have  a total of  88 quasi-simultaneous radio and X-ray observations of \gx\ during its hard state, covering a total of seven outbursts over 
a 15--year period. Our new measurements represent the largest sample for a stellar mass  
black hole, without any bias from distance uncertainties, over the largest flux variations and down to a level that could be close to quiescence, making \gx\ the reference source for comparison with other accreting sources (black holes, neutrons stars, white dwarfs and active galactic nuclei). 
Our results demonstrate a very strong and stable coupling  between  radio and X-ray emission, despite several outbursts of different nature and separated
by a period of quiescence. The radio and X-ray luminosity correlation of the form  L$_X \propto L_\mathrm{Rad}^{0.62 \pm 0.01}$ confirms the non-linear coupling between the jet  and the inner
accretion flow powers and better defines the standard correlation track in the radio-X-ray diagram for stellar mass black holes. We further note 
epochs of  deviations from the fit that significantly exceed  the measurement uncertainties, especially during the time of formation and destruction 
of the self absorbed compact jets. The jet luminosity could appear brighter (up to a factor 2) during the decay compared to the rise for a given X-ray 
luminosity, possibly related to the compact jets. We furthermore connect the radio/X-ray measurements to the near-infrared/X-ray  empirical correlation in \gx, further
demonstrating a coupled  correlation between these three frequency ranges. The level of radio emission would then be  tied to the near-infrared emission, possibly
by the  evolution of the broad-band properties of the jets.
We further incorporated our new data of  \gx\  in a more  global study of black hole candidates strongly supporting a scale invariance in the jet-accretion  coupling of accreting black holes,  
and confirms the existence of two populations of sources in the radio/X-ray diagram.

\end{abstract}


\section{Introduction}

Stellar mass black holes (BH) in accreting binaries undergo occasional outbursts with transitions between different spectral
states, usually defined  by their X-ray spectral and timing properties  \citep{McClintock06,Belloni10d}. In addition to the 
behaviour of  the inner accretion flow, outflows are   also crucial for the characterisation of these different spectral states \citep{Fender06b}, 
especially in light of the strong coupling between the jets and the disc.  Two main spectral states can be defined 
as follows: 1) the hard state is characterised by a  comptonised X-ray spectrum with a weak or absent (e.g. \citealt{Tomsick08}) thermal contribution 
from the accretion disc,  and the presence of powerful self-absorbed compact jets \citep{Corbel00, Dhawan00,Fender01, Stirling01}, 2) the soft state is mostly dominated 
by the thermal emission from the accretion disc and the absence of relativistic jets \citep{Fender99,Coriat11, Russell11b}. We further note the intermediate states  that
are associated with many changes in the X-ray spectral and timing properties   \citep{Belloni10d}  and the inflow - outflow coupling  
\citep{Corbel04b, Fender04b, Rodriguez08b, Fender09}. 

Weakly accreting BHs in the hard state have been the focus of many recent studies due to the
presence of several  emission processes that are very difficult to disentangle in this regime. 
Indeed, three main components are thought to dominate the energetic output of the system: 
the standard optically thick and geometrically thin accretion disc (eventually truncated in its inner part)  \citep{Shakura73}, 
the self-absorbed compact jets \citep{Corbel00,Fender01}, and an optically thin corona of hot electrons (\citealt{Sunyaev80}; see also \citealt{Xie12}) 
that may or may not be the base of the compact jets \citep[e.g.][]{Markoff05}. Emission from the companion
star is usually negligible for low mass X-ray binaries in outburst.

With the development of large campaign of multi-wavelength observations, several attempts have been 
conducted to model the spectral energy distribution (SED) of BHs in their regimes of faint X-ray emission
\citep[e.g.][]{McClintock01,Markoff03,Yuan05}. It turns out that optically thick synchrotron emission from the 
compact jets dominates the radio to near-infrared range in the hard state \citep{Corbel02a, Hynes03, Buxton04, Homan05, 
Russell06, Coriat09}. 
In the hard X rays,  inverse Comptonisation of the ambient photon field is the 
main emission process, with an active debate on whether it is thermal or non-thermal Comptonisation, or synchrotron 
self-Compton \citep[see for a review][]{McClintock06, Markoff10}. Optically thin synchrotron emission from the compact jets may 
also play a role at high energy   \citep{Markoff01, Homan05, Russell10,Laurent11a}. 

Another way to probe the emission properties of these BHs systems is to search for  empirical relations 
between various wavelengths. 
The tight non-linear power law correlation between radio and X-ray emission (of the form $L_{\rm Rad} \propto L_{\rm X}^{\sim 0.7}$
 in \gx\  \citep{Hannikainen98,Corbel00,Corbel03} demonstrated the strong coupling between the compact jets and 
the X-ray emitting media. This correlation originally brought to light the possibility of optically thin synchrotron X-ray emission 
 from the compact jets  of \gx\  \citep{Corbel03,Markoff03} or alternatively from a radiatively inefficient accretion 
flow (\citealt{Merloni03,Heinz04a}, but see also \citealt{Kording06}). \citet{Gallo03} extended this correlation study to a 
larger sample of BHs and proposed a universal correlation (still with an index of  $\sim$ 0.7) between the radio and 
X-ray luminosity for Galactic BHs.  The correlation observed in the hard state appears to be maintained down to the quiescent level of at least
two sources \citep[\vq, \a0\ and possibly also \gx :][]{Corbel03,Gallo03, Gallo06, Corbel08}. By taking into account the mass of the BHs, this correlation was extended 
to active galactic nuclei and led  to the definition of the fundamental plane of BH activity and the possibility of universal 
scaling laws for accreting BHs across all mass scales \citep{Merloni03,Falcke04,Kording06,Wang06,Gultekin09,Plotkin11}.  

However, it is important to note that the universal radio/X-ray correlation presented by  \citet{Gallo03} for BH binaries  is dominated by 
two sources (\gx\ from \citet{Corbel03} and \vq ) with few measurements from additional sources that are often close to
transition to the softer states. As the fundamental plane relies on the correlation for the Galactic BHs,  it is important 
to verify  how solid both the Galactic sample and the correlation are. \citet{Corbel08} revisited the correlation for \vq\  and confirmed the tight correlation from outburst 
down to quiescence with an index of $\sim$ 0.6 that could be  consistent with X rays emanating from synchrotron 
self-compton emission at the  base of the compact jets or Comptonisation from an inefficient accretion flow.  A
similarly tight correlation was also found between X-ray  and optical/near-infrared (hereafter OIR) emission in a sample of BH candidates (BHC)
and also in \gx\ over different outbursts \citep{Homan05, Russell06, Coriat09}. 
Alternatively to jet or ADAF models,  synchrotron emission from an hybrid electron distribution in a hot accretion flow has also   been invoked  to explain the OIR/X-ray correlation \citep{Veledina11}. 

In  recent years, significant efforts have been undertaken to observe  most new BHs 
transients in outburst at radio frequencies. This led to the discovery of several sources lying outside the scatter (hence  their names  ``outliers'') of the original radio/X-ray 
correlation (e.g. \xt , \citealt{Corbel04b}; \igr , \citealt{Rodriguez07}, \sw , \citealt{Cadolle07,Soleri10}). These outliers lie
significantly below the standard correlation \citep{Gallo12}. Using a large sample for the BH transient \1h, \citet{Coriat11} found a steeper correlation index of $\sim$ 1.4 for the outliers
(and possibly also other sources like \cx\ \citep{Zdziarski11b} and \grs\ \citep{Rushton10}). At fainter fluxes, 
these outliers (at least \1h, \citet{Jonker10,Coriat11} and possibly now also \maxisz\ and \xteds: \citealt{Ratti12, Jonker12}) may return to  the standard correlation. It remains unclear how these 
outliers are related to typical black  holes such as \gx\ or \vq . With a global study of \1h\ and other sources, \citet{Coriat11} 
proposed that  these outliers may represent a large subset of BHs radiating efficiently in the hard state. This is a change
of paradigm for BHs in the hard state,  usually believed to be in a regime of inefficient accretion (as also implied 
by the standard radio/X-ray correlation). Alternatively, a different  coupling between jet power and mass accretion rate \citep{Coriat11}  may 
also explain the steeper correlation index (see also \citealt{Soleri11}). 

In this paper, we review  the properties of the correlation  for \gx\ based on a very large sample of new data acquired during  
several  outbursts in the last decade.  Indeed, amongst the Galactic sources in the standard correlation (in fact dominated by
one outburst for \gx\ and \vq), none has been observed over several  outbursts.  The original work on \gx\ \citep{Corbel03} included 
data from the 1997 extended hard state  (e.g. \citealt{Harmon94,Corbel00}) and the decay from the 1998-99 outburst. The inclusion of two radio observations from the 
rise of the 2002-03 outburst pointed out  that \gx\ could follow another  track for different outbursts \citep{Nowak05}. 

Due to its recurrent outburst activity, \gx\ is an ideal target to probe the evolution of the correlation over distinct  outbursts. After the 
1998-99 outburst, \gx\ was found in quiescence in 2000 and 2001 \citep{Corbel03}. Since then, seven outbursts were observed: 2002-03 
\citep{Homan05,Belloni05b}, 2004-05 \citep{Belloni06}, 2006, 2007 \citep[e.g.][]{Tomsick08}, 2008, 2009,  2010-11 
\citep{Shidatsu11,CadolleBel11, Motta11,Rahoui12} and a temporary  reactivation in 2012 \citep{Lewis12,Maccarone12}. 
\gx\ is believed to harbour a BH with a mass $>$ 5.8~$M_{\odot}$ in a system with  a low mass star companion \citep{Hynes04}. \gx\ is located 
at a favoured  distance of 8 kpc \citep{Zdziarski04}.
In section 2, we present the radio and X-ray observations used in this analysis. The results for selected outbursts, as well as the re-evaluation of the 
radio/X-ray correlation using all data, are presented in Section 3. We then discuss these results in Section 4 in light of the multi-wavelength behaviour 
of \gx. Our conclusions are summarised in Section 5. 

\section{Observations and data analysis}

\subsection{ATCA observations}

All radio observations discussed in this paper were conducted with the Australia
Telescope Compact Array (ATCA)  located in Narrabri, New South Wales, Australia. 
The ATCA synthesis telescope is an east-west array consisting of six 22 m antennas.
It uses orthogonal linearly polarised feeds and records full Stokes parameters.  We 
carried out all  observations simultaneously at 4.8 GHz (6.3 cm) / 8.64 GHz (3.5 cm) and
in a limited cases also at 1.384 GHz (21.7 cm) / 2.496 GHz (12.0 cm). 
The ATCA was upgraded  in April 2009 with  the new  Compact Array Broadband Backend (CABB) system \citep{Wilson11}
resulting in significant improvement in sensitivity (an increase in bandwidth from 128 MHz to 2 GHz). This implied 
a small change in central frequencies and  the most recent observations were carried out at the frequencies of 5.5 and 9 GHz.  
Various array configurations have been used during these 
campaigns. All observations have been conducted by us, with the exception of  four observations (PI: M. Rupen)  in March 2004 (briefly mentioned 
in \citealt{MillerJ06a}). To compare with  the previous outbursts, we also used the original data of \gx\ from \citet{Corbel03} updated in
\citet{Nowak05}.  

The amplitude and band-pass calibrator was PKS~1934$-$638, and the antennas' gain and phase 
calibration, as well as the polarisation leakage, were usually derived from regular observations of the 
nearby (less than a degree away) calibrator PMN~1646$-$50.  The editing, calibration, Fourier 
transformation with multifrequency algorithms, deconvolution, and image analysis were performed using the {\tt MIRIAD} software package 
\citep{Sault98}.  
In short observations with  bad coverage of the u-v plane, flux densities were measured directly in the u-v plane 
with {\tt MIRIAD} task {\tt uvfit} and consequently checked afterwards with direct imaging of simulated datasets. 
We only present in this paper the data when \gx\ was in the hard state according to the X-ray behaviour (black squares in the
vertical track of the Hardness-Intensity Diagrams (Fig.~\ref{fig_hid}).

\subsection{RXTE observations}\label{Sect_Xray}

The {\em Rossi X-ray Timing Explorer} ({\em RXTE}) has conducted
(almost daily) monitoring   observations  of \gx\  during almost all outbursts
discussed here (see for example \citealt{Homan05,Belloni06,Motta11,Stiele11,CadolleBel11}), 
with an exposure time of typically 1 to 3 ks for each observations.  

We performed spectral analysis using data from the Proportional Counter Array \citep[PCA,][]{Jahoda06}
and the High Energy X-ray Timing Experiment  \citep[HEXTE,][]{Rothschild98} from
observations when \gx\ was in the hard state. 
The data were reduced using {\tt HEASOFT}  software package v6.11
following the standard steps described in the ({\em RXTE}) cookbook\footnote{http://heasarc.gsfc.nasa.gov/docs/xte/data\_analysis.html}.
We extracted PCA spectra from the top layer of the Proportional Counter Unit (PCU) 2
which is the best calibrated detector out of the five PCUs and the only one operational across all the observations we analysed. 
We  produced  the associated response matrix and added a systematic uncertainty of 0.5 \% 
to all spectral channels  to account for PCA calibration uncertainties. In addition, we used the background model (faint or bright) appropriate
to the brightness level of \gx\ to create background spectra.

For HEXTE, we produced a response matrix and applied the necessary dead-time correction. 
The HEXTE background is measured throughout the observation by alternating between the 
source and background fields every 32 s. The data from the background regions were then merged.
When possible we used data from both detector A and B to extract source and background spectra.
However, from December 2005, due to problems in the rocking motion of Cluster A, we extracted 
spectra from Cluster B only. HEXTE channels were grouped by 4 due to the low count rate 
in most of the observations. On Dec. 14$^\mathrm{th}$ 2009, Cluster B stopped rocking as well. From this date, 
we thus used only PCA data in our analysis.

\begin{figure}
\includegraphics[width=84mm]{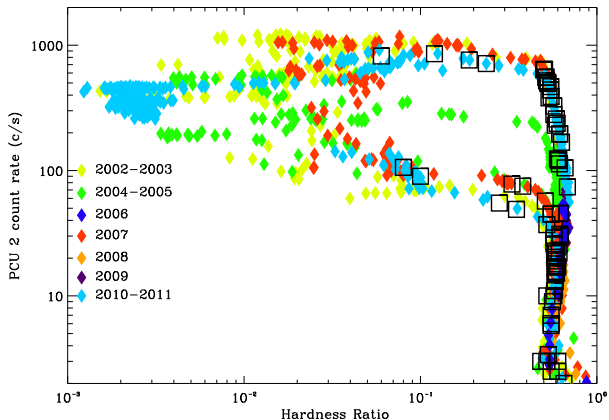}
\caption{{\em RXTE}/PCA hardness intensity diagrams (HID) covering all outbursts of \gx\ in the 2000-2011 period  (see end of section \ref{Sect_Xray} for definition of the X-ray hardness and intensity).  The squares highlight the epoch of our radio observations (not including the ones
conducted at very low flux).  Those on the vertical track are observations purely in the hard state, whereas  the few ones on the horizontal branches correspond to period of compact jet destruction in 2010, 
and formation in 
2007 and 2011 (see Section.~\ref{sect_indiv}).
}
\label{fig_hid}
\end{figure}

The X-ray fluxes in various energy bands were determined by fitting simultaneously 
the PCA and HEXTE spectra in \textsc{XSPEC} V12.7. 
We used a model consisting of a multi-color disc blackbody(\textit{ezdiskbb}, not required at low flux), a power-law with a high
energy cut-off when required (\textit{power $\times$ highecut}), and a Gaussian emission line (\textit{gaussian})
constrained in energy between 6 and 7 keV.  Interstellar absorption was taken into account using the \textit{phabs}
model with cross-sections from \citet{Balucinska-Church92} and abundances from \citet*{Wilms00}. 
We fixed the hydrogen column density to $N_\mathrm{H}$ =  6  $\times$ 10$^{21}$~cm$^{-2}$,  which is a typical value observed in 
\gx\ \citep[e.g.][]{Zdziarski04,CadolleBel11}.
A  multiplicative constant was added to the model to allow for differences in the 
overall normalisation between both instruments. All fluxes presented have been 
normalised to PCA and have been corrected for the interstellar absorption (its
effect is almost negligible for \gx\ above 3 keV).  The flux error bars were chosen 
to be the largest of the statistical error, or 3\%, which is a good estimate 
of the PCA  systematic uncertainty.  Furthermore, the X-ray fluxes from the original correlation \citep{Corbel03} have also been 
updated with the last version of the {\em RXTE} software and calibration. 
The level of Galactic Ridge emission \citep{Revnivtsev06,Revnivtsev07,Ebisawa08} 
has been estimated from long PCA observations taken when \gx\ was 
in quiescence (and therefore below PCA detection level, see \citealt{Coriat09} for details).
This level of  4 $\times$ 10$^{-12}$ \ergsc\ (3--9 keV band) has been subtracted from all X-ray fluxes.  Quantifying this level was
important for the discussion on the level of radio emission when \gx\ was detected at its faintest 
level of radio emission.

In addition, we constructed Hardness Intensity Diagrams (HIDs, see Fig.~\ref{fig_hid}) from  PCA data.
We produced background subtracted lightcurves binned at 16 s using PCA Standard 2 mode data from the PCU2 (all layers).
The lightcurves were divided into three energy bands, 2.5-6.1, 9.4-18.5 and 2.5-18.5 keV (according to the correspondence
between fixed energy channels and energy valid for the PCA gain epoch 5). We defined the hardness ratio (HR) as the ratio
of count rate in the bands 9.4-18.5 keV and 2.5-6.1 keV and the intensity as the count rate in the energy band 2.5-18.5 keV.

\subsection{Swift observations}

We used measurements made by the {\em Swift} XRT instrument to extend 
this study to low flux levels.  Our analysis included one 3.0 ks
observation made on 2009 October 30 that was previously reported by
\citet{Yen09}, five observations made in 2011 on 
April 13 (1.2 ks), April 27 (4.9 ks), May 11 
(4.8 ks), and September 9 (2.1 ks), and two observations made in 2012
on June 23 (0.15 ks) and June 26 (0.43 ks).  Although more {\em Swift} 
observations are available near these times, these are the observations 
that are closest in time to our radio observations.  

For all eight of these observations, XRT was operated in photon counting
(PC) mode, providing 2-dimensional imaging.  We analyzed the data using 
the HEASOFT software package.  For the 2009 and 2011 observations, we
reprocessed the data to produce cleaned event lists using the routine
{\ttfamily xrtpipeline}, and for the 2012 observations, we used the 
output of the standard pipeline processing.  We used {\ttfamily xselect}
to extract source spectra from a circular aperture with a 20 pixel
($47^{\prime\prime}$) radius and background spectra from an annular region
centered on the source.  After background subtraction, the 0.5--10 keV
count rates for the five observations range from $0.007\pm 0.002$ s$^{-1}$
to $0.14\pm 0.02$ s$^{-1}$, which are low enough for photon pile-up to 
be negligible.  We used the PC-mode response file
swxpc0to12s6\_20010101v012.rmf, and the routine {\ttfamily xrtmkarf} to 
produce the ancillary response (``arf'') files.  For the spectral fits 
described below, we combined the 2012 June 23 and June 26 spectra due 
to the short exposures and the fact that the radio observation was made 
on June 24th in between the two {\em Swift} observations.

To determine the X-ray flux, we used the XSPEC software package to fit
the energy spectra with a power-law model with interstellar absorption.
We used Cash statistics for the fits, which does not require binning of 
the spectra.  We fixed the column density to 
$N_{\rm H} = 6\times 10^{21}$ cm$^{-2}$, and the photon indexes obtained
for the non-quiescent observations ranged from $\Gamma = 1.5\pm 0.3$
to $2.0\pm 0.3$ (90\% confidence errors).  For the three quiescent
observations, the photon index was not well-constrained, and we fixed
it to $\Gamma = 2$.  The main goal of these spectral fits is to obtain 
the unabsorbed flux for each spectrum, and we accomplished this by using 
the model {\ttfamily pegpwrlw}, which has flux in an adjustable energy 
band as a free parameter, to represent the power-law.

\begin{figure}
\includegraphics[width=84mm]{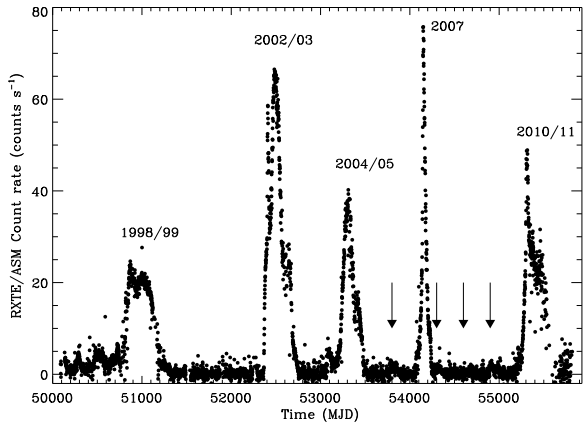}
\caption{{\it RXTE}/ASM soft X-ray (1.5-12 keV) light-curve of \gx\ for all RXTE lifetime (January 1996 to December 2011).  The major outbursts, that include a
full transition to the soft state, are indicated by their epoch.  In addition, the arrows point to additional fainter outbursts that are not visible 
with  ASM (but see the Swift light-curve on Figure \ref{fig_bat}) because \gx\ was only in the hard state. }
\label{fig_asm}
\end{figure}

\subsection{Notes on some individual observations } \label{sect_ind}

Almost all radio and X-ray observations were conducted  quasi simultaneously (less than a few hours). However, in only  very limited cases, we  
interpolated (usually with a second degree polynomial on a linear scale)  the monotonously rising/decaying X-ray flux  in order to get the most precise flux at the time of the radio observations. 

\begin{figure}
\includegraphics[width=84mm]{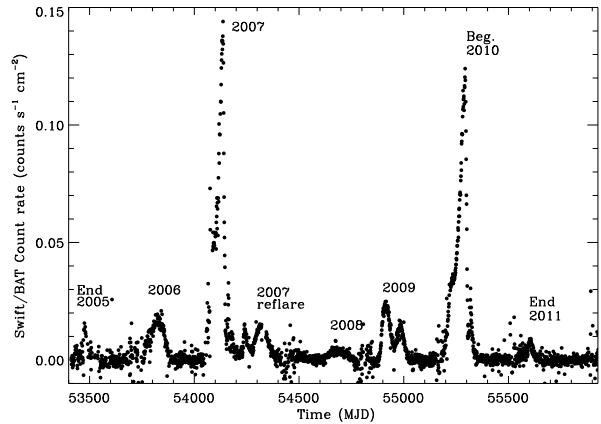}
\caption{{\it Swift}/BAT hard X-ray (15-50 keV) light-curve of \gx\  corresponding to Fig.~\ref{fig_asm} after {\it Swift} launch (February 2005 to December 2011). Outbursts are identified by their epoch. 
We note  that after the 2007 outburst, \gx\ brightened again (noted `2007 reflare') for few months  and stayed only in the hard state, as in 2006, 
2008 and 2009. }
\label{fig_bat}
\end{figure}

One single ATCA observation was conducted in  2006 on November 5  (MJD 54044.2) and it occurred well after the end of the faint 2006 outburst (from January to June 2006, 
see the {\it Swift}/BAT ligthcurve in Fig.~\ref{fig_bat}). Inspection of the {\it RXTE}/PCA light-curve (not shown) revealed that  the reported 2007 outburst already started, at least, as early
as 2006  November 12 (date of the first available PCA observation) with an 3--9 keV unabsorbed X-ray flux of 5.56 $\times$ 10$^{-11}$ \ergsc .
Thirteen RXTE/PCA observations occurred in a thirteen-day period and it showed  an X-ray flux that was steadily rising (see also \citealt{Swank06}). 
A  fit to the rising X-ray  flux leads to an extrapolated 3-9 keV flux of 2.2  $\times$ 10$^{-11}$ \ergsc\  (with a very conservative error of 30\%) for 2006, November 5 
(i.e 5 days earlier). 

During the 2008-2009 period, \gx\ is mostly detected by the hard X-ray all sky monitor (aka {\em Swift}/BAT), indicative of hard state activity only. 
For the observation on MJD 55002, the X-ray flux was taken from the {\em RXTE}/PCA Galactic Bulge Monitoring of Craig Markwardt\footnote{http://asd.gsfc.nasa.gov/Craig.Markwardt/galscan/main.html}.
On MJD 54469, the X-ray flux was estimated from an  average of the {\it RXTE}/PCA observations conducted 6 days before and after, at a period when \gx\ was not very variable. Finally,  
on 2011 May 11, we used the X-ray flux derived from a {\em Chandra} observations (to be discussed elsewhere) conducted at low flux.

Regarding the radio observations, we note that  in a few cases,  it was also necessary to take into account the presence of an additional component associated with the
interaction of  previously ejected material with the interstellar medium \citep[e.g.][]{Gallo04, Corbel10c}, similarly to what has been detected in \xte\ \citep{Corbel02b}. However, this 
did not concern any hard state measurements, but only the observations during the re-ignition of the compact jets in the 2007 soft to hard state transition. 
Finally, the first  ATCA  observations in 2002 was only made at 1.384 and 2.496  GHz and flux densities have been extrapolated to 8.6 GHz assuming a typical  \citep{Corbel03} radio 
spectral index $\alpha$ =+0.12, taking a definition of the flux density, $S_\nu$, as  $S_\nu \propto \nu^\alpha$ (consistent with the 1.384 to 2.496 GHz spectral index of 0.11 $\pm$ 0.11). Similarly, on 2002 April 5, the flux at 8.6 GHz is deduced from the detection at 4.8 GHz (no data usable at 8.6 GHz for this observations).  All our measurements representing the whole sample of \gx\ in the hard state are tabulated in Table 1.  
  
\begin{table*}
 \centering
 \begin{minipage}{166mm}
  \caption{Summary of data used for the radio/X-ray flux correlation in \gx}
  \begin{tabular}{@{}lccccccc@{}}
  \hline \hline 
   Calendar Date  &   MJD   & \multicolumn{2}{c}{Flux density}  & Radio  &  X-ray flux (3--9 keV) & X-ray  &   Comment  \\
         (UT)               &              &         4.8 or 5.5  GHz  		&	8.6 or 9.0 GHz 	 &           spectral            &  (in            unit of      10$^{-10}$                                      &  hardness   & \\
              		 & 		& 		\multicolumn{2}{c}{(mJy)}		 &		index		 & 		 \ergsc ) 						& 				&   \\
		           \hline

1997 Feb. 4	 & 50483.04	&       N.A.	                     & 9.10	$\pm$ 0.10  &  N.A.              & 	9.66	$\pm$ 0.34  	&  0.71  & C03,  N05  \\	
1997 Feb. 10	 & 50489.79	&       N.A.                     & 8.20	$\pm$ 0.20  &  N.A.              & 	8.48	$\pm$ 0.30	&  0.72  & C03,  N05 \\
1997 Feb. 17	 & 50496.81	&       N.A.  	          & 8.70	$\pm$ 0.20  &  N.A.              &	8.11	$\pm$ 0.29	&  0.72  & C03,  N05 \\
1999 Feb. 12 	 & 51221.96	&  6.34 $\pm$ 0.08    & 4.60	$\pm$ 0.08  &  --0.55 $\pm$ 0.04  &	4.39	$\pm$ 0.16	&  0.67  & C03,  N05, Na \\
1999 Mar. 3	 & 51240.71	&  6.07 $\pm$ 0.06    & 5.74	$\pm$ 0.06  &  --0.10 $\pm$ 0.03  &	4.35	$\pm$ 0.15	&  0.76  & C03,  N05,  Na \\
1999 Apr. 2	 & 51270.56	&  4.75 $\pm$ 0.06    & 5.10	$\pm$ 0.06  &  +0.12 $\pm$ 0.04  &	4.42	$\pm$ 0.16	&  0.53  & C03,  N05 \\
1999 Apr. 22	 & 51290.58	&  2.92 $\pm$ 0.06    & 3.20	$\pm$ 0.06  &  +0.16 $\pm$ 0.05  &	2.05	$\pm$ 0.07	&  0.55  & C03,  N05 \\
1999 May. 14	 & 51312.65	&  1.25 $\pm$ 0.05    & 1.44	$\pm$ 0.06  &  +0.24 $\pm$ 0.10  &	0.72	$\pm$ 0.03	&  0.52  & C03,  N05 \\
1999 Jun. 25	 & 51354.69	&  0.14 $\pm$ 0.03    & 0.24	$\pm$ 0.05  &  +0.92 $\pm$ 0.51   &	 (3.93 $\pm$ 0.28)$\times$10$^{-2}$	&  N.A   & C03 \\
1999 Jul. 7	 & 51366.56	&    $<$ 0.13               & 0.12	$\pm$ 0.04  &    N.A.           	 &	(1.53 $\pm$ 0.19)$\times$10$^{-2}$	&  N.A   & C03 \\
1999 Aug. 17	 & 51407.35	&      N.A.	                  & 0.27	$\pm$ 0.07  &    N.A.           	 &	(2.0 $\pm$ 0.3)$\times$10$^{-2}$&  N.A   & C03 \\
1999 Sep. 1	 & 51422.38	&      N.A.	                  & 0.32	$\pm$ 0.05  &    N.A.           	 &	(1.96 $\pm$ 0.10)$\times$10$^{-2}$ 	&  N.A   & C03 \\
1999 Sep. 15	 & $\sim$ 51436	&      N.A.	        &       $<$ 0.06	              &    N.A.            	&     (5.7 $\pm$ 1.0)$\times$10$^{-3}$	&  N.A   & C03 \\ 		
  
    & & & & & & &   \\

2002 Apr. 4	 & 52368.69	&    N.A.	                       & 5.95     $\pm$ 0.15  &  +0.11 $\pm$ 0.11 & 	18.62	$\pm$ 0.61	&  0.58 & N05, S.\ref{sect_ind}  \\    
2002 Apr. 5	 & 52369.66	&  6.77  $\pm$ 0.15  & 7.26	$\pm$ 0.20  &  N.A.             &	22.86	$\pm$ 0.73	&  0.57 &   S.\ref{sect_ind}   \\    
2002 Apr. 7	 & 52371.70	&  8.39  $\pm$ 0.06  & 8.27	$\pm$ 0.07  &  --0.02 $\pm$ 0.02 &	30.63	$\pm$ 0.96	&  0.54 & \\
2002 Apr. 18	 & 52382.86	&  13.19 $\pm$ 0.05  & 14.27	$\pm$ 0.05  &  +0.13 $\pm$ 0.01 &	53.86	$\pm$ 1.56	&  0.51 & N05 \\

    & & & & & & &   \\

2003 May 25	 & 52784.01	&   0.60 $\pm$ 0.05  & 0.77	$\pm$ 0.06  &  +0.42 $\pm$ 0.19 &	0.186	$\pm$ 0.030	&  0.60 &  \\ 
    & & & & & & &   \\

2004 Feb. 4	 & 53039.87	&  N.A.              & 0.70	$\pm$ 0.04  &   N.A.            &	0.145	$\pm$ 0.031	&  0.70 &  \\
2004 Feb. 13	 & 53048.89	&  1.27 $\pm$ 0.07   & 1.13	$\pm$ 0.08  & --0.20 $\pm$ 0.15  &	0.749	$\pm$ 0.027     &  0.52 & \\	
2004 Feb. 24	 & 53059.09	&  1.77 $\pm$ 0.15   & 1.84	$\pm$ 0.20  & +0.07 $\pm$ 0.23  &	2.36	$\pm$ 0.08	&  0.62 &  \\
2004 Mar. 16	 & 53080.79	&  4.75 $\pm$ 0.06   & 4.88	$\pm$ 0.06  & +0.10 $\pm$ 0.07  &	9.07	$\pm$ 0.32	&  0.61 &  \\
2004 Mar. 17	 & 53081.79	&  4.92 $\pm$ 0.05   & 4.84	$\pm$ 0.11  & --0.03 $\pm$ 0.05  &	9.55	$\pm$ 0.34	&  0.61 &  \\
2004 Mar. 18	 & 53082.79	&  5.15 $\pm$ 0.05   & 4.98	$\pm$ 0.11  & --0.05 $\pm$ 0.05  &	9.92	$\pm$ 0.35	&  0.61 &  \\
2004 Mar. 19	 & 53083.79	&  5.17 $\pm$ 0.04   & 5.20	$\pm$ 0.10  & +0.01 $\pm$ 0.04  &	10.06	$\pm$ 0.35	&  0.60 &  \\

    & & & & & & &   \\

2005 Apr. 21	 & 53481.73	&  4.35 $\pm$ 0.05   & 4.73	$\pm$ 0.05  & +0.14 $\pm$ 0.03 &	3.75	$\pm$ 0.13	&  0.55 &  \\
2005 Apr. 24	 & 53484.67	&  4.39 $\pm$ 0.06   & 4.23	$\pm$ 0.08  & --0.06 $\pm$ 0.04 &	2.77	$\pm$ 0.10 	&  0.57 &  \\
2005 Apr. 28	 & 53488.58	&  3.07 $\pm$ 0.10   & 3.46	$\pm$ 0.13  & +0.20 $\pm$ 0.08 &	1.85	$\pm$ 0.07	&  0.58 &  \\
2005 Apr. 29	 & 53489.65	&  2.88 $\pm$ 0.08   & 3.32	$\pm$ 0.10  & +0.24 $\pm$ 0.07 &	1.65	$\pm$ 0.06	&  0.58  &  \\
2005 Apr. 30	 & 53490.63	&  2.53 $\pm$ 0.05   & 2.94	$\pm$ 0.07  & +0.26 $\pm$ 0.05 &	1.57	$\pm$ 0.06      &  0.58 &\\	
2005 May 3	 & 53493.88	&  1.73 $\pm$ 0.10   & 1.92	$\pm$ 0.14  & +0.18 $\pm$ 0.16 &	1.17	$\pm$ 0.04	&  0.59 &  \\
2005 May 4	 & 53494.88	&  1.87 $\pm$ 0.08   & 1.99	$\pm$ 0.10  & +0.11 $\pm$ 0.11 &	1.25	$\pm$ 0.05	&  0.58 &  \\
2005 May 6	 & 53496.88	&  1.42 $\pm$ 0.09   & 1.69	$\pm$ 0.12  & +0.30 $\pm$ 0.16 &	0.96	$\pm$ 0.04	&  0.58 &  \\
2005 May 12	 & 53502.88	&  1.08 $\pm$ 0.12   & 1.00	$\pm$ 0.18  & --0.13 $\pm$ 0.36 &	0.73	$\pm$ 0.03	&  0.56 &  \\

    & & & & & & &   \\

2006 Nov. 5	 & 54044.17	&  0.37 $\pm$ 0.06  & 0.60	$\pm$ 0.07  & +0.82 $\pm$ 0.34 &	0.22	$\pm$ 0.08	&  N.A.  &  S.\ref{sect_ind}  \\
2007 Feb. 4a	 & 54135.10	&  19.5 $\pm$ 0.3   & 17.0	$\pm$ 1.0   & --0.23 $\pm$ 0.10 &	54.84	$\pm$ 1.92	&  0.50 &  S.\ref{sec_2007}\\
2007 Feb. 4b	 & 54135.77	&  21.9 $\pm$ 0.3   & 22.5	$\pm$ 0.3   & +0.05 $\pm$ 0.03 &	56.80	$\pm$ 1.99	&  0.50 & S.\ref{sec_2007}  \\

    & & & & & & &   \\

2007 May 31	 & 54251.74	& 4.34 $\pm$ 0.14   & 4.37	$\pm$ 0.09  & +0.01 $\pm$ 0.06 &	2.10	$\pm$ 0.08	& 0.57  &  \\   
2007 Jun. 6	 & 54257.74	& 2.63 $\pm$ 0.15   & 2.63	$\pm$ 0.18  & +0.00 $\pm$ 0.14  &	1.53	$\pm$ 0.06	& 0.59  &  \\
2007 Jun. 11	 & 54262.73	& 1.77 $\pm$ 0.18   & 2.01	$\pm$ 0.15  & +0.22 $\pm$ 0.21 &	1.28	$\pm$ 0.05	& 0.58  &  \\
2007 Jun. 25	 & 54276.71	& 1.69 $\pm$ 0.06   & 1.69	$\pm$ 0.05  & +0.00 $\pm$ 0.08 &	1.26	$\pm$ 0.05      & 0.59  & \\	
2007 Jun. 29	 & 54280.46	&  1.7 $\pm$ 0.2    & 2.0	$\pm$ 0.2   & +0.28 $\pm$ 0.26 &	1.33	$\pm$ 0.05      & 0.60  &\\	
2007 Jul. 4	 & 54285.52	&  1.7 $\pm$ 0.2    & 2.1	$\pm$ 0.2   & +0.36 $\pm$ 0.26 &	1.81	$\pm$ 0.07	& 0.61  &  \\
2007 Jul. 13	 & 54294.50	& 2.36 $\pm$ 0.05   & 2.66	$\pm$ 0.05  & +0.20 $\pm$ 0.05 &	2.50	$\pm$ 0.09      & 0.62  &\\	
2007 Aug. 22	 & 54334.50	& 3.30 $\pm$ 0.05   & 2.95	$\pm$ 0.07  & --0.19 $\pm$ 0.05 &	3.67	$\pm$ 0.13	& 0.62  &  \\
2007 Nov. 3	 & 54407.25	& 0.55 $\pm$ 0.06   & 0.80	$\pm$ 0.07  & +0.63 $\pm$ 0.24 &	0.25	$\pm$ 0.04	& 0.57  &  \\  
2007 Nov. 27	 & 54431.99	& 0.59 $\pm$ 0.07   & 0.48	$\pm$ 0.07  & --0.35 $\pm$ 0.32 &	0.25	$\pm$ 0.06	& 0.52  &  \\

    & & & & & & &   \\

2008 Jan. 4	 & 54469.05	& 0.70 $\pm$ 0.05   & 0.65	$\pm$ 0.05  & --0.13 $\pm$ 0.18 &	0.275	$\pm$ 0.063	&  N.A. & S.\ref{sect_ind}  \\
2008 Jun. 26	 & 54643.77	& 0.77 $\pm$ 0.08   & 1.16	$\pm$ 0.10  & +0.70 $\pm$ 0.23 &	0.714	$\pm$ 0.075	&  0.62 &  \\
2008 Jul. 5	 & 54652.70	& 1.08 $\pm$ 0.06   & 1.24	$\pm$ 0.07  & +0.24 $\pm$ 0.13 &	0.875   $\pm$ 0.092     &  0.62 &  \\	
2008 Jul. 16	 & 54663.70	& 1.21 $\pm$ 0.07   & 1.51	$\pm$ 0.06  & +0.38 $\pm$ 0.12 &	1.17	$\pm$ 0.12      &  0.64 & \\
2008 Aug. 18	 & 54696.60	& 1.10 $\pm$ 0.10   & 1.18	$\pm$ 0.10  & +0.12 $\pm$ 0.21 &	0.858	$\pm$ 0.090     &  0.62 & \\
2008 Sep. 29	 & 54738.02	& 0.69 $\pm$ 0.10   & 0.91	$\pm$ 0.10  & +0.47 $\pm$ 0.31 &	0.454   $\pm$ 0.49      &  0.57 & \\	
2008 Oct. 10 	 & 54749.37	& 0.86 $\pm$ 0.07   & 0.73	$\pm$ 0.10  & --0.28 $\pm$ 0.27 &	0.166   $\pm$ 0.021     &  0.76 & \\	

  \hline
\end{tabular}
\end{minipage}
\end{table*}

\begin{table*}
 \centering
 \begin{minipage}{166mm}
  \contcaption{Summary of data used for the radio/X-ray flux correlation in \gx}
  \begin{tabular}{@{}lccccccc@{}}
  \hline \hline
   Calendar Date  &   MJD   & \multicolumn{2}{c}{Flux density}  & Radio  & X-ray flux ( 3--9 keV) & X-ray  &   Comment  \\
         (UT)               &              &         4.8 or 5.5  GHz  		&	8.6 or 9.0 GHz 	 &           spectral            &  (in            unit of      10$^{-10}$                                      &  hardness   & \\
              		 & 		& 		\multicolumn{2}{c}{(mJy)}		 &		index		 & 		 \ergsc ) 						& 				&   \\

          \hline
  
  2009 Jun. 20	 & 55002.59	& 2.22 $\pm$ 0.03 	& 2.76	$\pm$ 0.03  & +0.44 $\pm$ 0.04 &	2.01	$\pm$ 0.51	&  0.65  & RXTE GBM\\
2009 Oct. 30	 & 55134.33	& \multicolumn{2}{c}{Combined upper limits: $<$ 0.05}&   N.A   &       (2.2 $\pm$ 0.8)$\times$10$^{-3}$ &  N.A.  & Swift/XRT   \\ 
  
      & & & & & & &   \\

 2010 Jan. 21	 & 55217.92	& 5.08 $\pm$ 0.04   & 5.05	$\pm$ 0.05  & --0.01 $\pm$ 0.03 &	6.02	$\pm$ 0.21	 & 0.68  & \\
2010 Feb. 13	 & 55240.01	& 6.17 $\pm$ 0.06   & 5.90	$\pm$ 0.10  & --0.09 $\pm$ 0.04 &	8.28	$\pm$ 0.29	 & 0.66  & \\
2010 Mar. 3	 & 55258.89	& 7.22 $\pm$ 0.10   & 7.30	$\pm$ 0.10  & +0.02 $\pm$ 0.04 &	14.06	$\pm$ 0.49	 & 0.63  & \\
2010 Mar. 6	 & 55261.89	& 9.02 $\pm$ 0.10   & 9.60	$\pm$ 0.05  & +0.13 $\pm$ 0.03 &	16.41	$\pm$ 0.58	 & 0.62  & \\
2010 Mar. 7	 & 55262.91	& 8.24 $\pm$ 0.05   & 8.05	$\pm$ 0.10  & --0.05 $\pm$ 0.03 &	16.49	$\pm$ 0.58	 & 0.62  & \\
2010 Mar. 14	 & 55269.80	& 10.18 $\pm$ 0.10  & 11.32	$\pm$ 0.10  & +0.22 $\pm$ 0.03 &	20.01	$\pm$ 0.70	 & 0.60  & \\
2010 Mar. 16	 & 55271.60	& 10.85 $\pm$ 0.10  & 12.04	$\pm$ 0.10  & +0.21 $\pm$ 0.03 &	21.91	$\pm$ 0.77	 & 0.59  & \\
2010 Mar. 21	 & 55276.82	& 13.76 $\pm$ 0.10  & 15.45	$\pm$ 0.06  & +0.24 $\pm$ 0.02 &	28.16	$\pm$ 0.99	 & 0.57  & \\
2010 Mar. 24	 & 55279.79	& 15.56 $\pm$ 0.05  & 18.59	$\pm$ 0.05  & +0.36 $\pm$ 0.01 &	33.09	$\pm$ 1.16	 & 0.55  & \\
2010 Mar. 28	 & 55283.73	& 19.48 $\pm$ 0.10  & 21.88	$\pm$ 0.10  & +0.24 $\pm$ 0.02 &	38.02	$\pm$ 1.33	 & 0.55  & \\
2010 Mar. 31	 & 55286.73	& 22.68 $\pm$ 0.05  & 25.94	$\pm$ 0.05  & +0.27 $\pm$ 0.01 &	40.57	$\pm$ 1.42	 & 0.55  & \\
2010 Apr. 2	 & 55288.91	& 21.95 $\pm$ 0.05  & 25.18	$\pm$ 0.10  & +0.28 $\pm$ 0.01 &	43.09	$\pm$ 1.51	 & 0.53  & \\
2010 Apr. 3	 & 55289.91	& 18.84 $\pm$ 0.10  & 21.11	$\pm$ 0.15  & +0.23 $\pm$ 0.02 &	44.36	$\pm$ 1.55	 & 0.53  & \\
2010 Apr. 4	 & 55290.82	& 21.60 $\pm$ 0.04  & 23.53	$\pm$ 0.05  & +0.34 $\pm$ 0.01 &	44.21	$\pm$ 1.55	 & 0.53  & \\
2010 Apr. 5	 & 55291.81	& 21.13 $\pm$ 0.04  & 24.69	$\pm$ 0.05  & +0.32 $\pm$ 0.01 &	45.25	$\pm$ 1.59	 & 0.52  & \\
2010 Apr. 6	 & 55292.81	& 21.30 $\pm$ 0.05  & 23.90	$\pm$ 0.06  & +0.24 $\pm$ 0.01 &	48.94	$\pm$ 1.71	 & 0.51  & \\
  
        & & & & & & &   \\

  2011 Feb. 13	 & 55605.81	& 4.45 $\pm$ 0.04  & 4.17	$\pm$ 0.05  & --0.13 $\pm$ 0.03 &	3.23	$\pm$ 0.11	 & 0.52  & \\
2011 Feb. 15	 & 55607.98	& 4.07 $\pm$ 0.04  & 3.87	$\pm$ 0.05  & --0.10 $\pm$ 0.03 &	2.59	$\pm$ 0.09      & 0.57  & \\
2011 Feb. 18	 & 55610.08	& 3.85 $\pm$ 0.10  & 3.98	$\pm$ 0.10  & +0.07 $\pm$ 0.07 &	2.03	$\pm$ 0.07      & 0.59  & \\
2011 Feb. 20	 & 55612.97	& 3.31 $\pm$ 0.05  & 3.84	$\pm$ 0.05  & +0.30 $\pm$ 0.04 &	1.41	$\pm$ 0.05 	 & 0.58  & \\
2011 Feb. 24	 & 55616.69	& 2.54 $\pm$ 0.04  & 2.95	$\pm$ 0.05  & +0.30 $\pm$ 0.05 &	0.966	$\pm$ 0.035      & 0.59  & \\
2011 Feb. 27	 & 55619.73	& 1.75 $\pm$ 0.10  & 2.42	$\pm$ 0.08  & +0.66 $\pm$ 0.13 &	0.760	$\pm$ 0.028	 & 0.57  & \\  
2011 Mar. 3	 & 55623.66	& 1.32 $\pm$ 0.04  & 1.64	$\pm$ 0.05  & +0.44 $\pm$ 0.09 &	0.610	$\pm$ 0.023	 & 0.55  & \\
2011 Mar. 7	 & 55627.10	& 1.11 $\pm$ 0.05  & 1.26	$\pm$ 0.10  & +0.25 $\pm$ 0.19 &	0.500	$\pm$ 0.019	 & 0.55  & \\
2011 Mar. 9	 & 55629.66	& 1.13 $\pm$ 0.05  & 1.38	$\pm$ 0.08  & +0.41 $\pm$ 0.15 &	0.460	$\pm$ 0.018	 & 0.55  & \\
2011 Mar. 20	 & 55640.04	& 0.63 $\pm$ 0.03  & 0.74	$\pm$ 0.04  & +0.33 $\pm$ 0.15 &	0.240	$\pm$ 0.042      & 0.48  & \\	
2011 Mar. 22	 & 55641.95	& 0.62 $\pm$ 0.05  & 0.76	$\pm$ 0.05  & +0.41 $\pm$ 0.19 &	0.16	$\pm$ 0.04	 & 0.55  & \\
2011 Apr. 15	 & 55666.89	& 0.30 $\pm$ 0.02  & 0.40	$\pm$ 0.03  & +0.58 $\pm$ 0.20 &	(3.1 $\pm$ 1.0)$\times$10$^{-2}$	 & N.A.  & Swift/XRT \\
2011 Apr. 27	 & 55678.84	& 0.88 $\pm$ 0.10  & 0.39	$\pm$ 0.10  & --1.03 $\pm$ 0.54 &	(1.4 $\pm$ 0.4)$\times$10$^{-2}$	 & N.A.  & Swift/XRT\\
2011 May 12	 & 55693.89	& \multicolumn{2}{c}{Combined upper limits: $<$ 0.060}  &  N.A. &   (1.78 $\pm$ 0.10)$\times$10$^{-3}$ 	 & N.A.  & Chandra/ACIS  \\  
2011 Sep. 9	 & 55813.33	& \multicolumn{2}{c}{Combined upper limits: $<$ 0.054}  &  N.A. &       (2.2 $\pm$ 1.0)$\times$10$^{-3}$	 & N.A.  & Swift/XRT  \\  
								
      & & & & & & &   \\

2012 Jun. 24	 & 56102.75	& 0.30 $\pm$ 0.04  & 0.35	$\pm$ 0.05  & +0.31 $\pm$ 0.39 &	(5.4 $\pm$ 1.8)$\times$10$^{-2}$  &  N.A. & Swift/XRT  \\

  \hline
\end{tabular}

NOTE: C03: see also  \citet{Corbel03}. N05: see also  \citet{Nowak05}. The calendar dates correspond to the the mid-point of the radio observations.  Upper limits are given at a 3$\sigma$ confidence level.
The Galactic ridge emission has been subtracted from the quoted  unabsorbed 3--9 keV X-ray flux.  See end of section \ref{Sect_Xray} for the definition of the X-ray hardness.  GBM:  X-ray flux from {\em RXTE}/PCA Galactic Bulge Monitoring of Craig Markwardt. Swift/XRT: X-ray flux from Swift/XRT. Chandra/ACIS: X-ray flux from Chandra/ACIS. 
S.x.x: See complementary information in Section x.x.  Na: Data points from original C03 sample, but these two points with optically thin radio spectra are representative of the re-ignition of the compact jets during the soft to hard state transition (e.g. see section \ref{sect_indiv}). 
\end{minipage}
\end{table*}

\section{Results}

\subsection{A long series of outbursts in a decade}

In the last decade, \gx\ has been one of the most active accreting BHs with many outbursts (Figs.~\ref{fig_asm}   and \ref{fig_bat}). The work on the radio/X-ray flux correlation by
\citet{Corbel03} used data from 1997,  when \gx\ was in a period of extended and persistent hard state, and data from the decay of the 1998/99 
outburst. Four major outbursts have been observed since 2000: 2002/03, 2004/05, 2007 and 2010/11. In addition, the analysis of the  {\it Swift}/BAT
light-curve reveals weaker outbursts in 2006, 2008 and 2009, when \gx\ was observed only in the hard state. After the decay of
the 2007 outburst, \gx\ brightened again (defined as ``2007 reflare" in Fig.~\ref{fig_bat})  but stayed only in the hard state.
We have plotted in Fig.~\ref{fig_hid} the full hardness intensity diagram (HID). 
For 1998/99, very  few RXTE observations were conducted and therefore these HIDs are not represented. For a general and recent overview of the  activity, see e.g. \citet{Dunn08,Droulans10,Wu10, Motta11,Stiele11,CadolleBel11,Rahoui12,Buxton12}.

\subsection{\gx\ close to quiescence}\label{sect_quiesc}

Following the CABB upgrade (April 2009) of the  ATCA, three of our radio observations have been conducted  when \gx\ was in a very faint 
state of X-ray emission (2009 October 30, 2011 May 12 and 2011 September 9), prior and after the 2010/11 outburst. Although  these radio observations did not result in a positive
detection when considered individually (Table 1), we decided to combine them together as the simultaneous Swift observations gave a similar 
X-ray flux (an unabsorbed  3--9 keV flux of $\sim$ 2 $\times$ 10$^{-13}$ \ergsc\ with an error of the order of $\sim$ 10\%). 

No detection  was achieved in the combined 9 GHz dataset with an rms noise level of 17 $\mu$Jy, whereas the combined 5.5 GHz provided a 
detection of \gx\ at the level of 73 $\pm$ 16 $\mu$Jy (noise level in the image of 10  $\mu$Jy). All together, the 5.5 and 9 GHz data  (but with a significantly higher noise level at 9 GHz)
gives a detection of 45 $\pm$ 16 $\mu$Jy.  The choice
of the final radio flux in this very faint state depends of the assumed spectral shape: 45 $\pm$ 16 $\mu$Jy  if the spectrum is flat or  73 $\pm$ 16 $\mu$Jy at 
5.5 GHz in case of steep spectrum (which will translate to 54 $\mu$Jy at 9 GHz for a --0.6 spectral index). For the rest of the paper, we will assume a
level of 45 $\pm$ 16 $\mu$Jy,  which is consistent with the expected spectral shape of compact jets.  

Whether or not it represents the true  level of \gx\ in quiescence, it is however interesting to note that its X-ray emission has been 
hovering  around the same value for all X-ray observations conducted between outbursts  during the last decade \citep[e.g.][]{Corbel03,Gallo03b,Corbel05,Yen09}.
The faintest upper limit recently reported by \citet{Maccarone12} is only a factor $\sim$ 2.5 lower than the above X-ray flux, meaning that  
it should  at least represent the level of X-ray emission, to within a factor 2,  between the major outbursts. It  may  also be close to the level in quiescence, indicating that
\gx\ may be at the upper end of the quiescent BH luminosities \citep{Corbel06}. 

\subsection{The overall flux correlations}

\begin{figure}
\includegraphics[width=84mm]{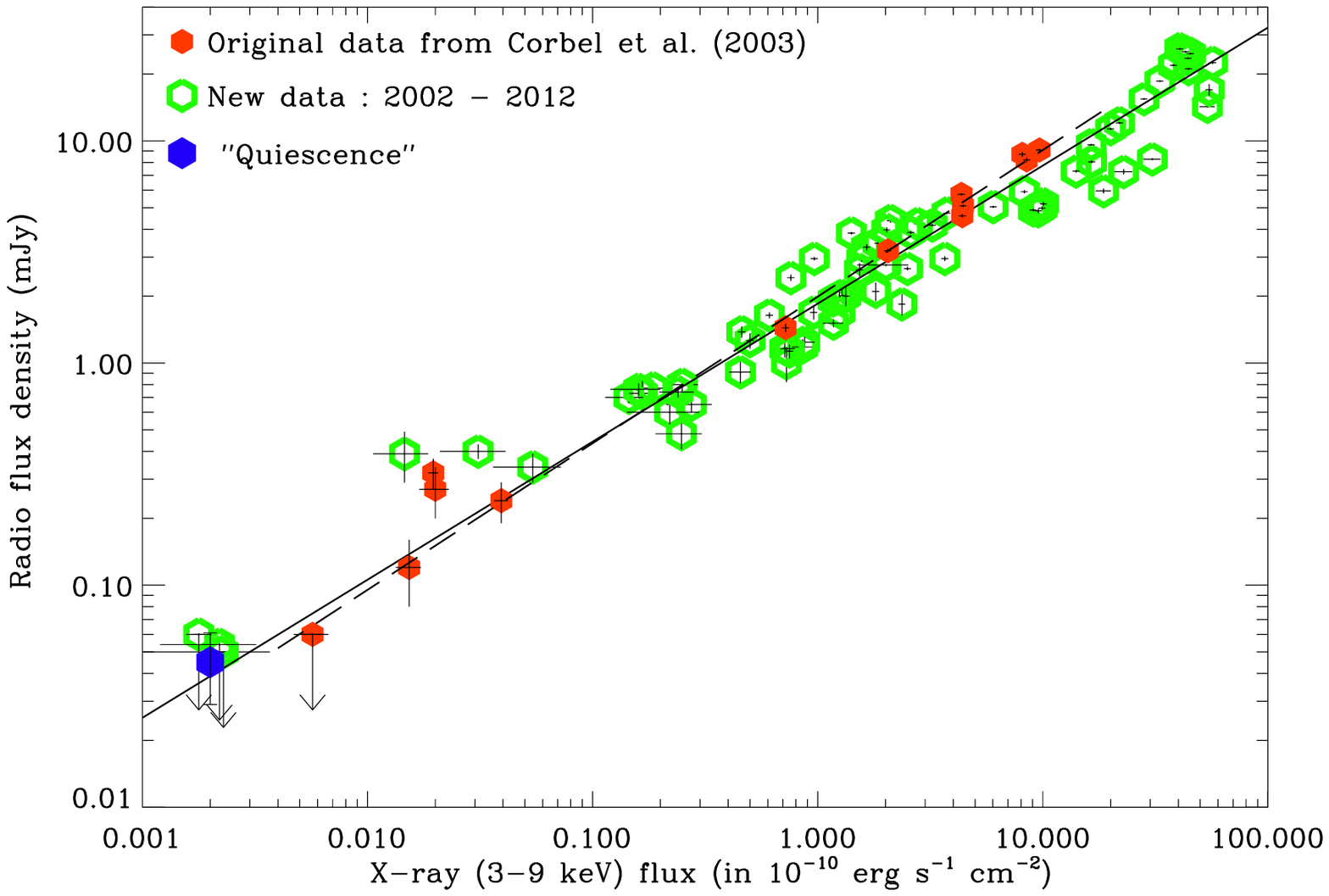}\
\caption{9 GHz radio emission  from \gx\ in the hard state versus the un-absorbed 3-9 keV flux for the new data-points (empty green hexagons) 
and the old data-points (filled red hexagons) from the original correlation \citep{Corbel03}. The dashed line illustrates the 
original radio/X-ray correlation in \gx\ from the 1997-1999 period  \citep{Corbel03,Nowak05}. The full  line corresponds to a fit to
the new whole sample with a function of the form $F_\mathrm{Rad}$ $\propto$ F$_\mathrm{X}^b$ (with $ b = 0.62  \pm  0.01 $). 
The blue point  corresponds to the level of \gx\ close to quiescence (see Section \ref{sect_quiesc}).}
\label{fig_correlall1}
\end{figure}

\subsubsection{ The Radio/X-ray flux correlation}\label{sect_global}

The new radio and X-ray measurements  of \gx ,   taken during the last ten years,  allow us to complement the original study of \citet{Corbel03} by adding measurements  
from seven additional outbursts, including four new outbursts that encompass a full X-ray states cycle (i.e. a full path in a given HID). In addition, it also samples the 
initial and final hard states  for a given outburst. 
These measurements are the most complete data-set available to study the evolution of the radio/X-ray correlation  of a BH over several outbursts  
(see also  \citealt{Coriat11} for \1h).  This new dataset  represents an increase of the original sample of \citet{Corbel03} by almost a factor 7 (88 observations instead of 13), 
plus 12 observations close to state transitions. 

\begin{figure}
\includegraphics[width=84mm]{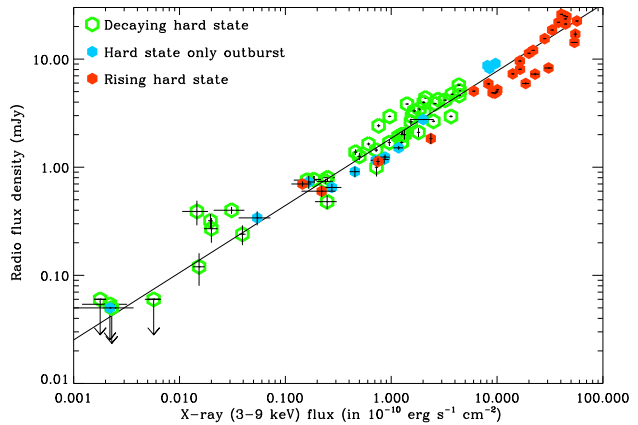}
\caption{Same as Fig. ~\ref{fig_correlall1} but with the data-points separated by the type of hard state within outburst: rising hard state (filled red hexagons), 
decaying hard state  (empty green hexagons)  and  outburst in persistent hard state only (filled blue hexagons). The full  line corresponds to the 
fit to the new whole sample.}
\label{fig_correlall2}
\end{figure}

The whole sample is represented in Fig. ~\ref{fig_correlall1}, where we separate the new data-points (in green) from the original sample 
(red points) of \citet{Corbel03}. In Fig.~\ref{fig_correlall2}, we 
divide the full sample according to the rise and  the decay of the outbursts, as well as highlighting those outbursts when \gx\ was found 
in the hard state only (see also Fig. ~\ref{fig_correlall3} for the data associated to each individual outburst).

The new data-points lie very close to the original correlation, illustrated by the red points in Fig.~\ref{fig_correlall1} and the updated fit (dashed line) to the old data (taking into 
account the upgrade of the {\it RXTE}/PCA softwares and calibration). In addition, they extend the correlation further to lower and higher fluxes. The new sample 
covers almost  five decades in X-ray flux and confirms that the correlation between radio and X-ray fluxes in the hard state of \gx\ is 
indeed very strong. The correlation is remarkably stable over a period of 15 years, even though several full outbursts (separated by off states) occurred  during this period, 
which might have affected  the structure/geometry  of the accretion flow.

The larger sample allow us to characterize more precisely  the functional form of this correlation. If we note, $F_\mathrm{rad}$, the 
radio flux density at 9 (or 8.6)  GHz (in mJy) and, $F_X$, the unabsorbed 3-9 keV  X-ray flux (in unit of 10$^{-10}$ \ergsc ); a fit (not using the upper limits) 
to all hard state data-points with a function of  the form: 
\begin{equation}
F_\mathrm{rad} =a~F^b_X,  
\end{equation} 
led to $a = 1.85    \pm  0.02 $ and $ b = 0.62  \pm  0.01 $  (to compare with  $a =  1.99  \pm  
0.04  $ and  $ b = 0.66 \pm  0.02  $ for the 1997-99 period taking into account the {\it RXTE}  using up-to-date instrument response files).  All reported  errors for the fitted functions 
are provided at the 90 \% confidence  level. The detection of \gx\  at low flux (Section \ref{sect_quiesc}), represented 
by a blue point in Fig.~\ref{fig_correlall1},  is also consistent with this function, meaning that the global correlation could be maintained down to quiescence 
if the level we measured is really representative of this low flux state.   The derived index of the correlation for \gx\ is fully consistent with the value (0.63 $\pm$ 0.03)
re-obtained recently by \citet{Gallo12} for the upper track of the correlation (their sample does not include our new \gx\ data).

By looking at  Fig. ~\ref{fig_correlall1}, which compares the original and new measurements, we note a significant 
dispersion (beside measurement errors) of the data-points around the fitted function. 
This dispersion is also well illustrated by Fig.~\ref{fig_correlall2}, where data-points are separated according to the phase (rise or decay) of an outburst
 We recall that data-points from the original correlation mostly 
originated from the decay of the 1998-99 outburst. The new data allow us to sample the rising hard state (not observed before with ATCA), 
that is known to reach a  higher luminosity than the decaying hard state \citep{Maccarone03}.  As expected,  higher 
levels of radio and X-ray emission  are observed during the rising hard state. 
The higher dispersion above $\sim$10$^{-10}$ \ergsc\ is partly related to differences between the rise and decay of outbursts. Indeed, 
the decaying hard state tends to be more radio bright for a given X-ray luminosity (or more X-ray faint for a given radio luminosity). 
This will be further discussed in Section  \ref{sect_indiv} on several individual outbursts.\\

\subsubsection{Comparison with the OIR/X-ray flux correlation}\label{sect_oir}

The radio emission from \gx\ in the hard state originates from the self-absorbed compact jets \citep{Corbel00}. With an inverted radio spectrum (i.e. a positive 
spectral index, $\alpha$, and a definition of the flux density, $S_\nu$, as  $S_\nu \propto \nu^\alpha$ ), the synchrotron emission from the compact  jets also reaches the
infrared range  \citep{Corbel02a, Homan05, Russell06, Gandhi11}, implying that a strong correlation  between the near-infrared  and X-ray fluxes should also exist. This has indeed 
been demonstrated by \citet{Coriat09} for the 2002-2007 period (see also \citealt{Homan05}). 

\begin{figure}
\includegraphics[width=84mm]{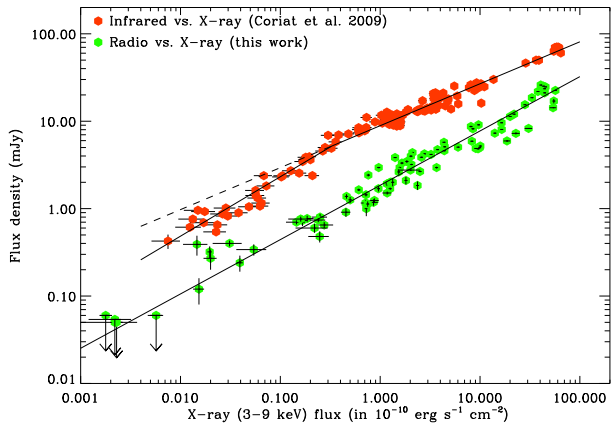}
\caption{9 GHz radio emission  (green points)  and  H band (1.6 $\mu$m) near-infrared  emission (red points from Coriat et al. 2009) from \gx\ versus the un-absorbed 3-9 keV flux. The solid lines 
represent the fitting function according to the sample. For the infrared data, the dashed line illustrates the break at low X-ray flux.
}
\label{fig_correlradir}
\end{figure}

We plot in Fig.~\ref{fig_correlradir} our new  radio/X-ray sample together with the entire infrared/X-ray sample from \citet{Coriat09}. The total magnitudes have been converted to flux units 
using a total extinction of  A$_\mathrm{V}$ = 3.7 mag. \citep[see][ for more details]{Coriat09,Buxton12}. 
 With a radio to infrared 
spectral index that is usually positive, it is not surprising to find brighter infrared emission than  radio emission for a given X-ray flux (we note that the accretion disc may
eventually contribute slightly to the near-infrared emission at low flux).  As demonstrated 
by \citet{Coriat09}, a broken power-law was necessary to fit the near-infrared/X-ray sample ($\chi^2$  = 584 for 130  degrees of  freedom), and here 
the radio/X-ray sample was fitted with a single power-law  ($\chi^2$  = 6001 for 94 degrees of freedom). We can not constrain the presence of a break 
in the radio/X-ray flux correlation  due to the low  number of radio detections at low flux density. The resulting reduced $\chi^2$ further indicate that intrinsic 
variability is also present along the fitted functions. The variance of the residuals (data over model) is 0.026 for the near-infrared/X-ray sample and 0.14 for 
the radio/X-ray sample. This highlights (see also Fig.~\ref{fig_correlradir})  a larger dispersion for the radio data along the fitted function compared to the
near-infrared sample.

\subsection{Focus on some individual outbursts}\label{sect_indiv}

We now turn to a more detailed discussion of three outbursts that have been well covered in radio and in particular the early phase of the decaying hard state when the compact
jets are building up. It is important to understand the nature of the larger dispersion (not of statistical origin)  of the radio data, 
especially in light of possible difference in the normalisation  between the rising and decaying hard  states \citep[e.g.][for \xte]{Russell07b}. 

\subsubsection{The 2004-2005 outburst}\label{sec_2004}

In early 2004,  \gx\ began a new outburst \citep{Buxton04b}, which ended around mid-2005. The radio coverage of the
hard states was  much better than during the previous outbursts with 8 observations during the  rise  and 9 observations during the decay  phase. 
The initial hard state was covered from 2004 February 3, i.e. before the X-ray detection of the outburst \citep{Smith04}
and ended on March 19 (close to the top of the right branch of the HID).
It then spent almost five months in this part of  the HID before starting a transition to a softer state around mid-August 2004 \citep{Buxton04c,Homan04a,Belloni06}. 
According to \citet{Buxton05},  \gx\ underwent a transition back to the hard state around  mid-April 2005. When the first ATCA observation  was conducted on 2005 April 21, 
the radio emission from \gx\ was already decaying with an inverted spectrum  (spectral index of 0.14  $\pm$ 0.03), suggesting the full 
reactivation of  the compact jets.

\begin{figure}
\includegraphics[width=84mm]{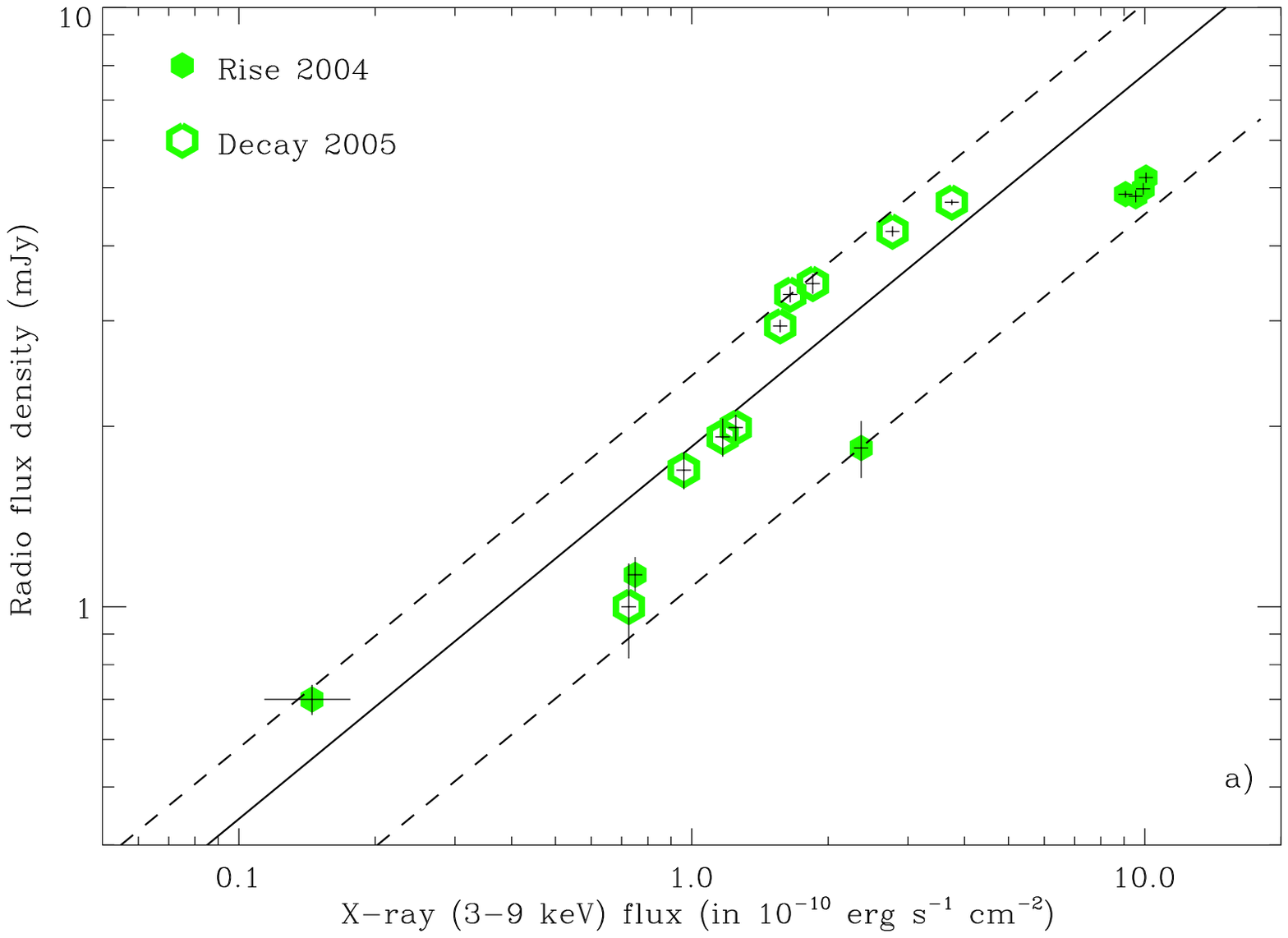}
\includegraphics[width=84mm]{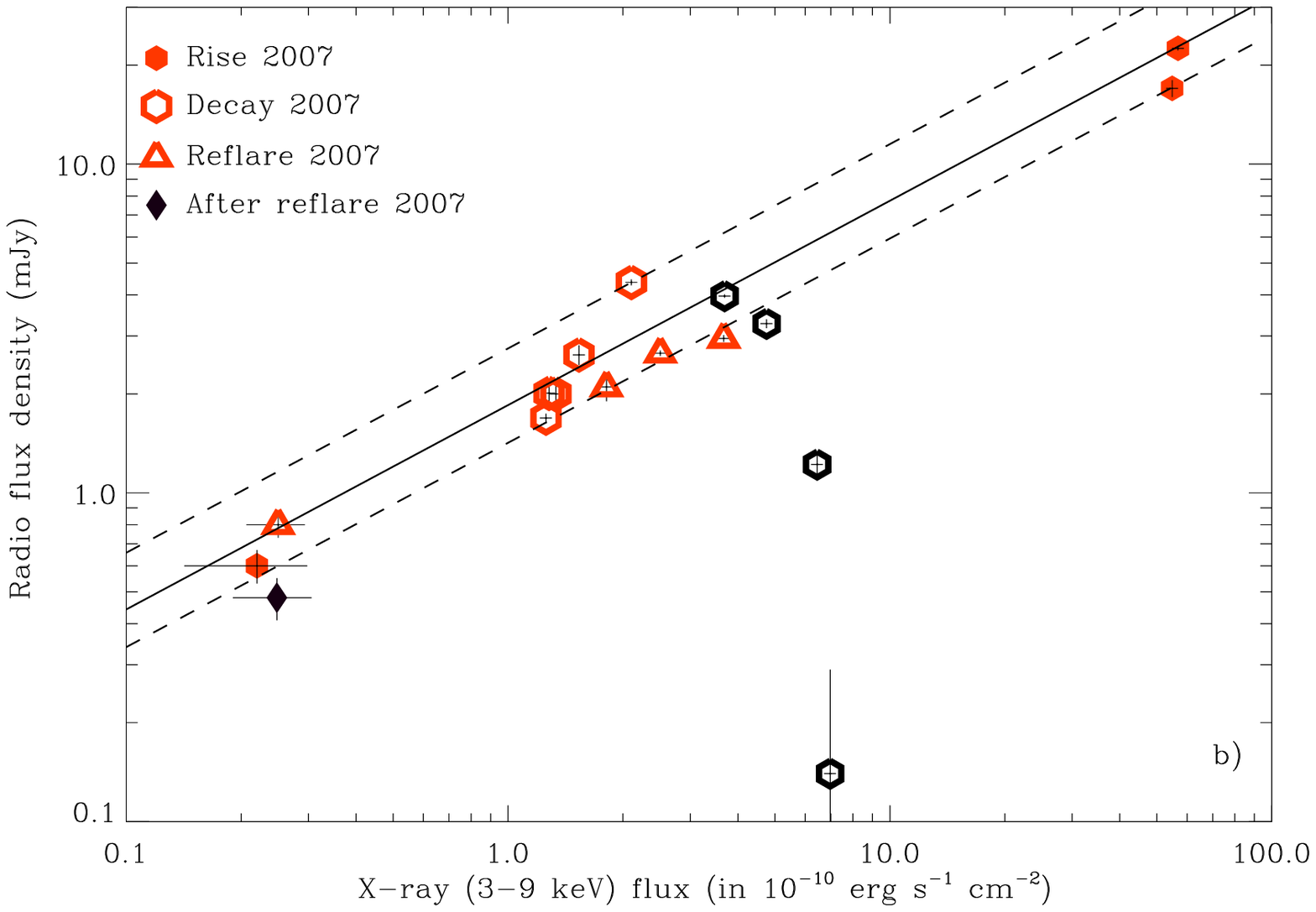}
\includegraphics[width=84mm]{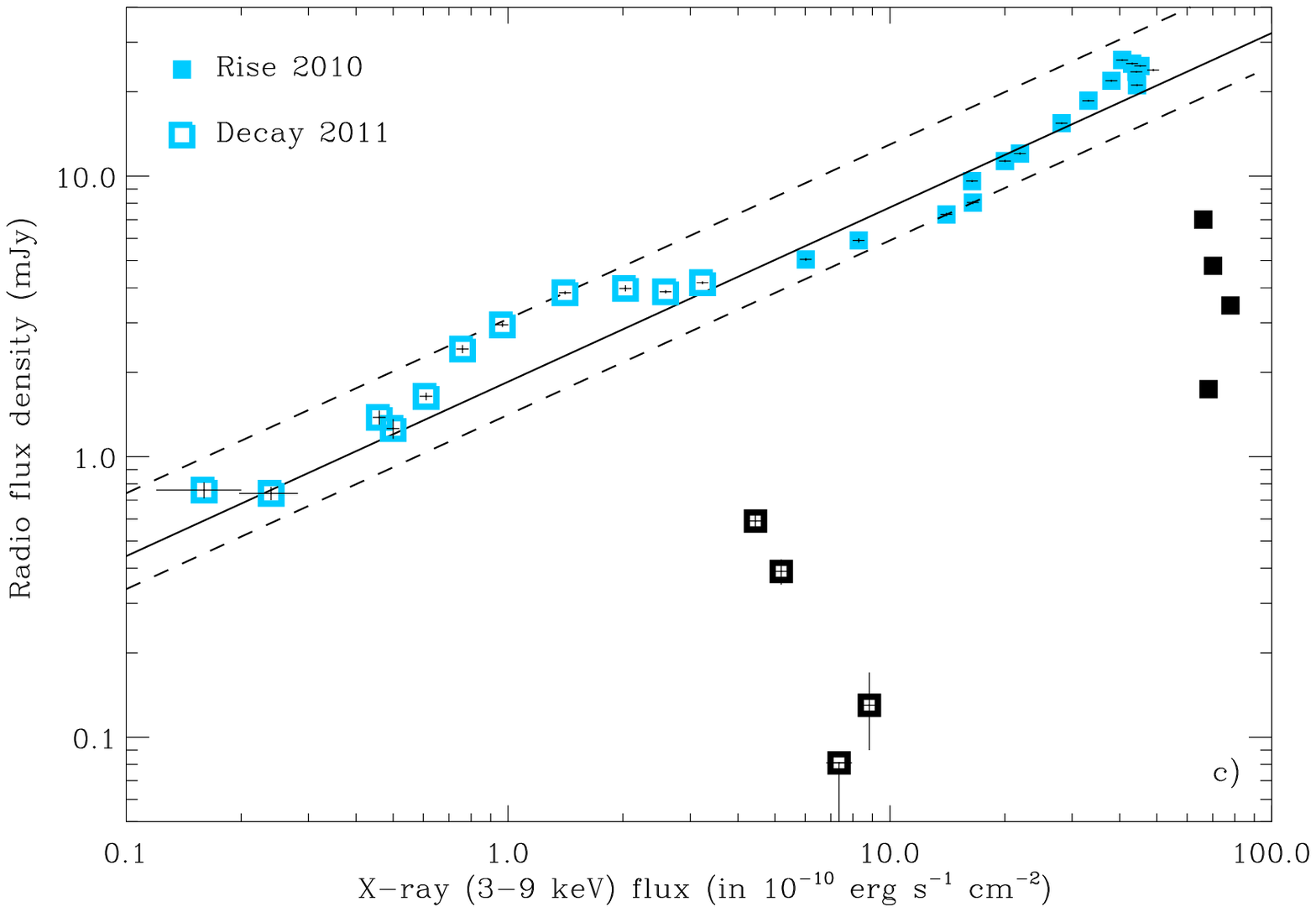}
\caption{9 (or 8.6) GHz radio flux density  from \gx\ versus the un-absorbed 3-9 keV flux for the X-ray rise (filled symbols) and the X-ray decay (empty symbols) for the three
outbursts of \gx\ with good radio coverage during both  the X-ray rising and decaying hard states.  The epoch of outburst are highlighted in the top left corner of each panel. The 
time of the reflare in 2007 is also indicated in the middle panel.  The solid line illustrates the  fit to the whole 1997-2012 sample with a function of the form 
$F_\mathrm{Rad}$ $\propto$ F$_\mathrm{X}^b$ (as discussed in Section \ref{sect_global}). The dashed lines illustrate the variation in normalisation of 
the fitted function needed to accommodate the observed flux variations between the rise and  the decay with the slope fixed to the value obtained for the whole sample. 
The black hexagons and squares  in the middle and bottom panels illustrate the turn-off (filled) and turn-on (empty) of the compact jets. Note that the axis ranges are not identical in all figures.}
\label{fig_correl_separate}
\end{figure}

Figure \ref{fig_correl_separate} (a)  illustrates the location of the radio and X-ray measurements for the 2004/05 outburst compared to the fit for the whole 
1997-2012 sample. In early February  2004, at the faintest level of radio/X-ray emission, \gx\  was close to  
 the global correlation. As the outburst evolved during the rising hard state, \gx\ moved significantly below this track. 
For the decay,  a slightly different behaviour can be noticed once \gx\ moved back to the hard state. In the initial part of the 2005 decay,  
at the brightest level of radio/X-ray  emission, \gx\ was slightly above  the  global correlation. 
The difference between the rise and decay tends to lower at fainter X-ray fluxes. 
For this decay, the radio spectra are always consistent with optically thick synchrotron  emission, which characterises the compact jets.
The correlations are therefore not affected by potential ``relic" optically thin synchrotron  emission from  previously  ejected  plasma that could 
interact with the ambient medium \citep{Corbel10c}.

If we compare the brightest measurements during the  2004 rise and the  2005 decay of the outburst, we find  that the decaying hard state is 
significantly more radio bright  for a given X-ray flux (or less X-ray bright for a given radio flux density) compared to the rising hard state. We find a  difference of a factor 
$\sim$ 2.25 in normalisation if we fix the slope to the value obtained for the full sample (see dashed lines in Fig.~\ref{fig_correl_separate}).
 The difference is particularly important   if we consider the soft to hard state transition, once the compact jets are being rebuilt. This slow evolution  of the correlation is clearly one of the
reasons for the scatter in the radio/X-ray correlation. This  behaviour will be discussed in more detail in Section  \ref{sect_discu}.  

\subsubsection{The 2007 outburst}\label{sec_2007}

Compared to the previous  outbursts, the one that occurred in 2007 was rather short. Indeed, with a reactivation in November 2006 (\citealt{Swank06} and 
Section \ref{sect_ind}), the transition to  the soft state occurred in March 2007 \citep{Motta09}, with a transition back to the hard state after mid-May \citep{Kalemci05b}.
The  X-ray  flux decreased up to mid-June, when a new hard state reflare was observed (see Fig.~\ref{fig_bat}). It was associated with  a re-brightening in radio, infrared and X-ray \citep{Tomsick08,Coriat09}. 

One radio observation was conducted at very faint fluxes close to the detectable onset of the outburst  (Section \ref{sect_ind}), whereas two other radio observations occurred in the 
brightest portion of the rising hard state (on the same day  -- February 4 -- but separated by  12 hours), i.e. close to  the top of 
the right branch in the HID.  
 These three observations are plotted (filled red hexagon) in Fig. \ref{fig_correl_separate} (b). We can 
simply note that they are  consistent with the global radio/X-ray correlation. One may possibly see, at high X-ray flux, a  potentially  rapid  increase in the radio emission from \gx\  (see also the 2002 and 2010 rises in Fig. \ref{fig_correlall3}) that could possibly be related to the onset of  a radio flare (e.g. \citealt{Gallo04}). However, those flares usually occur at much lower X-ray hardness \citep{Fender99}  (it is also likely the case in \gx, e.g. see section
\ref{sec_2010}) and is probably not a valid explanation for the observed increase here.

The decay in 2007 has also been well sampled in term of radio observations with the ATCA. Indeed, for the
first time, the onset of the compacts jets has been observed during the transition from the intermediate to the hard state 
(see also  \citealt{Corbel00} and the 2011 decay). The radio spectra are initially consistent  with optically thin synchrotron. But as the radio emission 
rapidly increases (see black hexagones in Fig.~\ref{fig_correl_separate} b),  it becomes and stays optically thick. As the focus of this paper is the long term study of 
the radio X-ray  correlation in \gx, we will discuss this aspect together with the behaviour at different wavelengths in a forthcoming paper.
We note that, during the formation of the compact jets, the 3--9 keV X-ray emission from \gx\ steadily declined as usual in the hard state.  Once the radio emission  is fully 
consistent with the compact jets, the radio emission also gradually  decreases. 

In terms of the radio/X-ray correlation, the  data points during the decay (Fig.~\ref{fig_correl_separate} (b)) are indicated by the open red hexagons, whereas the open red triangles 
highlight the reflare during this decay. 
Despite only five  radio observations for this standard decay (as the re-flare occurred soon after the hard state transition), 
we again note that for a given X-ray flux, \gx\ appears more radio bright during the decay compared to the rise. A change  in the normalisation of a  factor  at least 1.94 
 would be needed if we want to move the lower track (associated to the rising hard state)  to accommodate the brightest radio observation in the decay
(keeping the slope fixed to the value obtained for the full sample).

Regarding the re-flare portion of the 2007 activity, we can simply note that all data-points are strongly correlated with a slope 
 consistent with the global correlation.  A radio observation conducted on November 27  after the end of the 2007 re-flare (black diamond) is also 
indicated. It corresponds to a period of very weak residual hard state activity from \gx. It is again globally consistent with the overall correlation.

\subsubsection{The 2010-2011 outburst}\label{sec_2010}

A new full outburst began in 2010 as first noted by the MAXI  X-ray monitoring \citep{Yamaoka10}. After an initial hard state \citep{Tomsick10b, Corbel10b}, a transition 
to the soft state was reported by \citet{Belloni10b}. The transition back to the hard state occurred around mid-January 2011 \citep{MunozDarias11,Russell11a}. 
Compared to previous outbursts, the rising hard state has been well covered with 20 radio observations conducted almost  up to  the transition to the soft state. The very early part 
of the rise has, however,  not been sampled this time. A similar number of observations was conducted for the decaying hard state, 
meaning that it is the best sampled outburst with the most sensitive (due to CABB) ATCA radio observations. 

\begin{figure*}
\includegraphics[width=168mm]{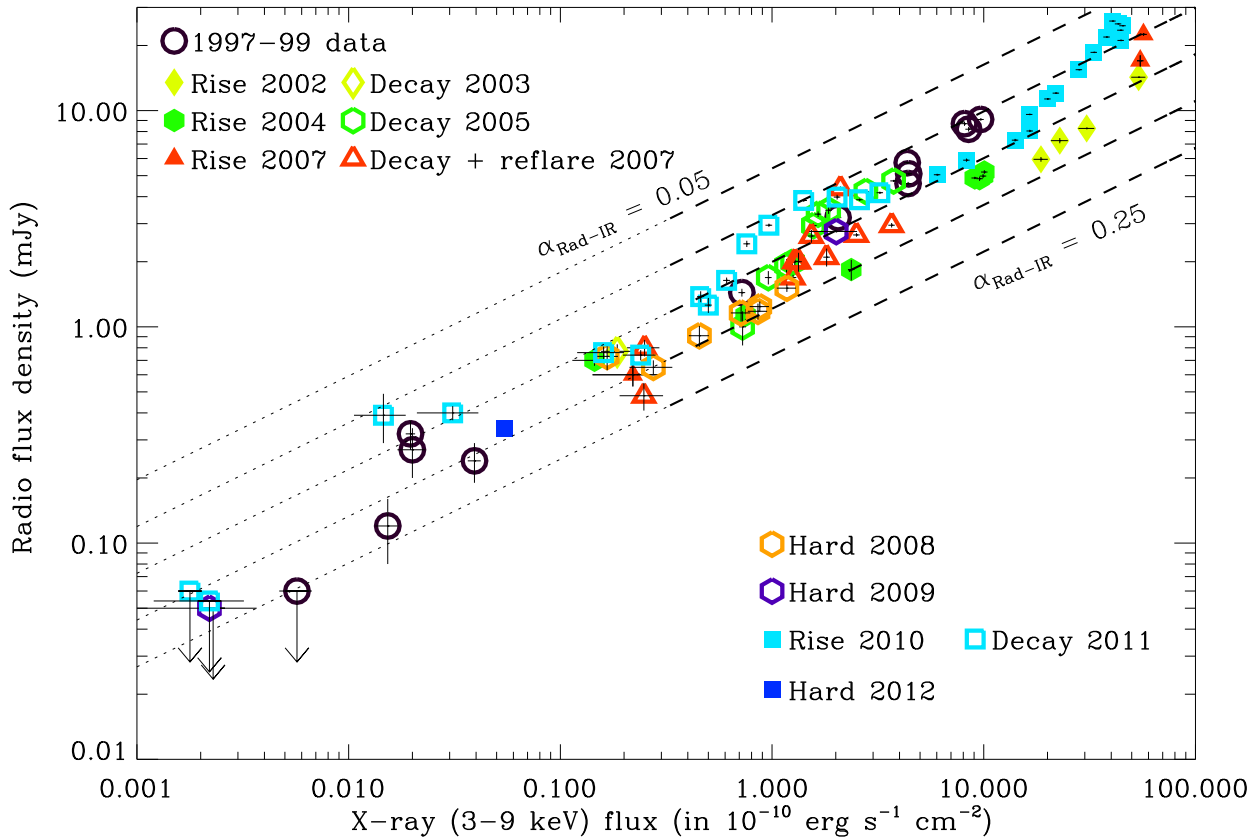}
\caption{9 GHz radio emission  from \gx\ versus the un-absorbed 3-9 keV flux for the whole 1997 - 2012 period.  The different outbursts are highlighted, as well 
as some individual portions (e.g. rise or decay) if necessary. The lines correspond to the estimated (not a fit) radio flux densities based on the relation of \citet{Coriat09} using the 
observed near-infrared/X-ray correlation (valid for X-ray fluxes above 3 $\times$ 10$^{-11}$ \ergsc, a region highlighted by the thicker lines), only assuming a radio to near-infrared spectral index, $\alpha_{Rad-IR}$, 
ranging from 0.25 to 0.05 with a step of 0.05. The extrapolation to lower fluxes is indicated by the  dotted lines (but see text for cautions). }
\label{fig_correlall3}
\end{figure*}

Inspection of the radio and X-ray measurements in Fig.~\ref{fig_correl_separate} (c) shows that they are rather well correlated and consistent with the global
correlation discussed in the previous subsections. However, as the outburst evolves in the rising hard state,  we note that  \gx\  seems to get  more radio bright 
(or less X-ray bright) for a given X-ray flux,  possibly  implying a steeper correlation  at higher flux. 
This steepening of the correlation at higher flux (very close to the peak flux in hard state) may possibly be a recurrent 
phenomena, as similar behaviour seems to be present also in 2002 (Fig.~\ref{fig_correlall3}) and 2007 (see Section \ref{sec_2007}). 
At the end of the hard state, the radio emission starts to be quenched as observed previously  \citep{Fender99, Corbel00}.  It also indicates that the major radio flare
in \gx\ occurs at a hardness significantly below the one defined by the hard state.

The transition back to the hard state occurred at lower X-ray flux  defining an hysteresis cycle \citep{Maccarone03}. Similarly to 2007, the radio emission is initially optically
thin (open black squares in Fig.~\ref{fig_correl_separate} c). The radio emission then becomes  and remains optically thick (a signature of a 
full re-building of the compact jets) once \gx\ gets close 
to the  global radio/X-ray correlation. Again, similarly to the decay in 2005 and 2007, it is initially above the standard correlation and as the X-ray emission decays, \gx\ 
is observed to move back to the global correlation. 
In the very bright parts of the hard state (and close to the state transition), significant thermal emission from the accretion disc could contribute to the X-ray spectrum. If we
only keep the contribution from the power-law,  the points close to the transition (black squares in Fig.~ \ref{fig_correl_separate} c) move obviously towards lower
X-ray fluxes, but not back to the global correlation (as it may be the case in \cx ; \citealt{Zdziarski11b}).

If we adjust a line  to fit the two extremes in the radio and X-ray diagram (lower track for the rising hard state in 2010 and upper
track for the decaying hard state in 2011) but with a slope fixed to the global correlation, we need a factor 2.19 (identical to 2004-2005) to accommodate the observed 
variations. Again, the larger dispersion along the global correlation occurs around the time of compact jet formation.  

\section{Discussion}\label{sect_discu}

We have presented a set of 100 quasi-simultaneous radio and X-ray observations of \gx, including 12 
observations close to state transitions.   
 This constitutes the  largest sample of measurements over a broad range of fluxes and also  over the longest time period for a black 
hole in the hard state.  In addition, this long term study  of \gx\ allows us to present  the evolution of the radio/X-ray flux correlations over several 
 outbursts separated by off states for a given BH without any bias from distance uncertainties.  
 This makes \gx\ the reference source for comparison with other accreting sources (BHs, neutrons stars, white dwarfs and AGN). The main characteristics from the new set of data can be summarised as follows: 
\begin{itemize}
\item  Despite numerous outbursts of different nature (full or hard state only) during a 15-year period, the overall correlation between the radio and X-ray fluxes is
strongly maintained. It is observed over five decades in X-ray flux and three decades in radio flux. By fitting all measurements  (radio flux density at 9 GHz in 
mJy, $F_\mathrm{rad}$,   and   unabsorbed 3-9 keV  X-ray flux in unit of 10$^{-10}$ \ergsc , $F_X$) with a  function: $F_\mathrm{rad} =a~F^b_X$,    we obtained  $a = 1.85   \pm  0.02  $ and $ b = 0.62  \pm  0.01 $. 
The faintest levels  of X-ray and radio emission, possibly \gx\ in quiescence, are also consistent with the global correlation. 
\item  We note a significant dispersion around the fitting function in the radio/X-ray diagram that appears more pronounced than for the IR/X-ray correlation (from \citealt{Coriat09}).
However, the dispersion of the radio measurements  does not occurr randomly. The strongest deviation seems to occur in 
 periods when the compact jets are  turning on, during the soft to hard state transition (see 2005, 2007 and 2011 decaying hard states, see  Fig.~\ref{fig_correl_separate}). 
 Additional deviations may exist at higher flux  during the rising hard state. 
\item During the period of compact jet formation in the decaying hard state ( $<$  3$\%$ \ledd\ for a 6 M$_\odot$ BH at 8 kpc),  we note that  \gx\ seems to be more radio bright (or less X-ray loud) than during the corresponding rising hard state, resulting in a difference of normalisation of the order of 2, that seems  constant over the outburst history of \gx. 
\end{itemize}

\subsection{A coupled radio,  near-infrared and X-ray correlation}\label{sect_fdlir}

 In Corbel et al. (in prep.),  we  study in detail the formation of the compact jets during  the soft to hard state transitions. 
We show that the compact jets  first turn on at radio frequencies and then  in the OIR (a conclusion that has also recently been  reached by \citealt{MillerJones12b}). 
However, the timescale to reach the peak is much longer in the near-infrared compared to the radio domain (see for example also Fig. 5 in \citealt{Coriat09}).
The consequence is that the radio emission quickly decays,  whereas the near-infrared emission is still rising, when the 3--9 keV X-ray flux 
is steadily decreasing. This implies a significant  evolution of the radio to near-infrared spectra  (equivalent to major changes in the radio to near-infrared  flux ratio, 
e.g. see \citealt{Coriat09}) during the soft to hard state transition. 

The dispersion in the near-infrared / X-ray flux  correlation is smaller than in a comparison between radio and X-ray emissions (Section \ref{sect_oir}). This is consistent 
with the standard picture where the NIR emission zone is located close to the base of the jet and is therefore more tightly connected to the X-ray emitting flow than the 
termination of the compact jets where the radio emission arises.
In that  case, the exact  location of \gx\  in the radio / X-ray flux diagram would then be influenced by the radio to near-infrared spectral index (or similarly by the radio to near-infrared flux ratio), 
i.e. the internal energetic balance of the compact jets. 

Indeed, in a generic model of self absorbed compact jets \citep{Blandford79}, the jet emission is  characterized by a flat to  inverted spectrum from the radio up to a turnover frequency,
that is estimated to be in the mid to near-infrared range  for Galactic BHs \citep{Corbel02a, Coriat09, Gandhi11, Rahoui11}. 
In this optically thick regime, the radio, $F_\mathrm{Rad}$,  to near-infrared, $F_\mathrm{Near-IR}$,  flux densities can be written as :  
$F_\mathrm{Rad}$  $\propto  F_\mathrm{Near-IR} ~\nu^{\alpha_{\mathrm{Rad-IR}}}$, with $\alpha_{\mathrm{Rad-IR}}$  the radio to near-infrared spectral index  and $\nu$ the frequency. 
Using the relation of \citet{Coriat09} between the near-infrared flux densities and the 3--9 keV X-ray fluxes, F$_\mathrm{x}$, of  $F_\mathrm{Near-IR}$ = k F$_\mathrm{x}^{b_2}$ 
(with $b_2$ = 0.48 $\pm$ 0.01 for the portion of the correlation at high flux), we obtain the following simplified relation: 
\begin{center}
\begin{equation}
F_\mathrm{Rad}  = \mathrm{k}   \left( \frac{\nu_{\mathrm{Rad}}}{\nu_{\mathrm{Near-IR}}} \right)^{\alpha_{\mathrm{Rad-IR}}} F_\mathrm{x}^{b_2},
\label{eq_fr}
\end{equation}
\end{center}
with $\nu_{\mathrm{Rad}}$ and $\nu_{\mathrm{Near-IR}}$ the frequencies corresponding to the radio and near-infrared data (k  is adapted  from Table 1 of \citealt{Coriat09}). 

However, a constant radio to near-infrared spectrum (i.e. $\alpha_{Rad-IR}$=constant) should normally lead to a dependence between radio and X-ray emission of the same 
kind as observed between infrared and X-ray emission,  i.e. $F_\mathrm{Rad}$ $\propto$  F$_\mathrm{x}^{b_2}$ (with $b_2$ = 0.48 $\pm$ 0.01, \citealt{Coriat09}). 
This is not what is exactly observed as we found  $F_\mathrm{Rad}$ $\propto$  F$_\mathrm{x}^{b}$,  with  $ b = 0.62  \pm  0.01 $  (Section \ref{sect_global}). 
But,  equation (\ref{eq_fr}) also highlights  an extra dependance of the normalisation  with the radio to near-infrared spectral index $\alpha_{Rad-IR}$, that could influence the
exact location of  \gx\ in the radio/X-ray flux diagram (beside the X-ray flux) and could affect the fitted correlation index b.  

As a complementary comparison, we plot in Fig.~\ref{fig_correlall3}  the simultaneous radio/X-ray measurement for the various outbursts of \gx . 
The different lines  (not a fit) correspond  to the predicted radio fluxes  based on the measured   infrared/X-ray correlation (equation \ref{eq_fr}) from \citet{Coriat09} (we use the 
functional form with no break that is valid for 3-9 keV flux above $\sim$ 3 $\times$ 10$^{-11}$  \ergsc ),  assuming a radio 
to near-infrared spectral index $\alpha_{\mathrm{Rad-IR}}$  in the  range of 0.05 to 0.25, which are typical values for the hard state. 
These small changes in the radio to near-infrared spectral index are sufficient to  explain the overall dispersion  in the radio and X-ray flux correlation, especially  
the larger dispersion  above  $\sim$10$^{-10}$ \ergsc (because it corresponds to larger variation in $\alpha_{\mathrm{Rad-IR}}$, see Corbel et al. in prep.). 
Such an evolution of the radio to near-infrared spectral index could simply be due to deviation from the simple Blandford--K\"onigl model used for  compact jets
(such a different jet geometry and/or different electron energy distributions, e.g. \citealt{Heinz06,Kaiser06}) or the efficiency of particles acceleration in the jets (Corbel et al. in prep.). 

However, one needs to  be cautious with equation \ref{eq_fr}, which assumes that the near-infrared observing band lies on the optically thick part of the jet spectrum. While
this should be fine at high flux (although not really tested), this may not be the case at lower flux (see \citealt{Coriat09} for more details). Furthermore, the near-infrared emission 
of \gx\ at high flux in the hard state is dominated by the compact jets (e.g. the large drop in OIR emission during the hard to soft state transitions, \citet{Homan05, Coriat09,Buxton12}), but the thermal emission from the accretion disc may contribute at low flux in the total near-infrared 
emission from \gx. Therefore, the thick lines in Fig.~\ref{fig_correlall3} should be a guide to highlight a possible link between radio, near-infrared and X-ray emission,
but the full understanding of the radio to near-infrared SED is beyond the scope of this paper.

\begin{figure*}
\centerline{\includegraphics[width=170mm, angle=0]{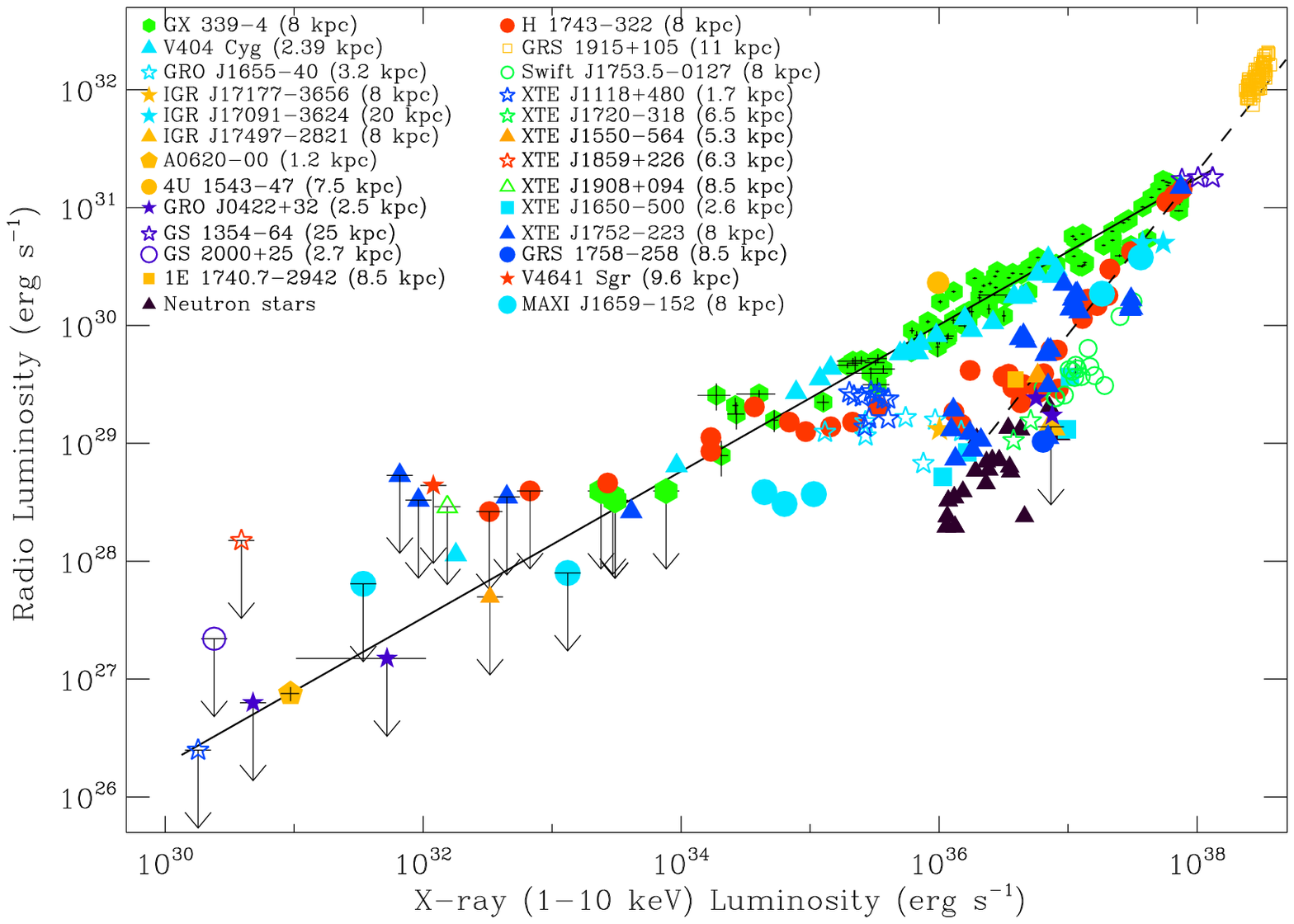}}
\caption{Radio  and X-ray (1-10 keV) luminosities for Galactic accreting binary BHs  in the hard and quiescence states (see text for references). 
It illustrates  the standard correlation (defined by sources such as \gx\ or \vq\ with index $\sim$ 0.6) and the new correlation for the outliers (defined 
by e.g.  \1h\ or \sw\ with index $\sim$ 1.4 ). The solid line illustrates the  fit to the whole 1997-2012 sample of \gx\ (as discussed in Section \ref{sect_global})
with an extrapolation to the quiescence state of BHs. The dashed line corresponds to the fit to the data for \1h, one of the representatives 
for the outliers \citep{Coriat11}. Upper limits are plotted at the 3$\sigma$ confidence level. For a few sources, their distances are unknown.  }
\label{fig_correlAll}
\end{figure*}

\subsection{The ``Universal'' radio/X-ray flux correlation}

\subsubsection{The ``standard'' track}

The new sample of \gx\   provides a new opportunity to compare the radio and X-ray emission  of  
Galactic accreting binary BHs and neutron stars. In Figure~\ref{fig_correlAll}, we report 
our new data  of \gx\ along with the measurements from published  studies representing a total of 24 BHCs. This includes the very few sources with very good radio
and X-ray coverage (i.e. \vq\ from \citet{Gallo03} and \citet{Corbel08}; \1h\ from \citet{Coriat11}; \grs\ from \citet{Rushton10},  \sw\ from \citet{Soleri10},
\maxisz\ from  \citet{Ratti12} and  \xteds\ from \citet{Jonker12} and Brocksopp et al. to be submitted) and the very few measurements from additional sources reported  in more global studies (e.g. \citet{Gallo03, Calvelo10, Fender10, 
Coriat11b}) or individual reports (e.g. IGR~J17177$-$3656 from \citealt{Paizis11}; IGR~J17091$-$3624 from \citealt{Rodriguez11}). To avoid confusion in the figure, we do not 
plot the large group of measurements related to \cx . Two neutron star systems are also reported for comparison: Aql~X$-$1 from \citet{Tudose09, MillerJones10} and 4U~1728$-$34 from  \cite{Migliari06b}. 
In addition, we also report the quiescence levels of a sample of BH binaries \citep{Calvelo10, MillerJones11}. 
All X-ray luminosities have been converted to the 1-10 keV band and we used the updated distances from \citet[][and references therein]{Fender10}.  For some sources, especially towards the Galactic
bulge, the distance may be unknown and is assumed to be at 8 kpc. 

With the  distances of \gx\  \citep{Zdziarski04} and \vq\ \citep[updated in ][]{MillerJones09}, these two sources now share the same track  in the radio/X-ray diagram. The updated fit  to 
the whole  \gx\  sample is  represented  on Figure~\ref{fig_correlAll}  by the solid line.
The quiescent level of \a0\   \citep{Gallo06} is remarkably consistent with an extrapolation of the \gx\ fit to
lower fluxes. However, the exact track of \a0\ during its outburst is unknown as no radio observation was conducted during its hard state
at  that epoch (see also \citealt{Yuan09}).  All other upper limits from quiescent BHs does not contradict the extrapolation of the correlation defined by \gx . 
The extension of the radio/X-ray correlation with index $\sim$ 0.6, as determined by the new data of \gx, down to the faintest levels of emission from Galactic black
hole binaries, implies that the standard correlation could be stable (despite minor deviations within an outburst) over more than 8 decades 
in X-rays  and 6 decades in radio luminosities.  Furthermore, no break in the correlation  is observed at lower luminosity with the available data (as expected 
in some theoretical model , e.g. \citealt{Yuan05b}), but this could occur at even lower luminosity \citep{Yuan09}. This means that more sensitive observations  (e.g. with the upgraded JVLA/ATCA and/or the 
new forthcoming SKA pathfinders)  of quiescent BHs (or BHs in their  decaying hard state) would  be needed to better constrain the slope of the correlation at low luminosities,  as it  currently relies 
only on the  quiescence level of \a0, \vq\ and possibly now also \gx. 

\subsubsection{The ``outliers'' track}\label{sect_outliers}

A growing number of sources (originally  \xt\ \citep{Corbel04b},  \igr\  \citep{Rodriguez07}, \sw\  \citep{Cadolle07,Soleri10})  are now 
detected significantly below the standard correlation. The number of these outliers is rapidly increasing  (see  Figure~\ref{fig_correlAll}). At  a 1--10 keV luminosity 
 above 4 $\times$ 10$^{36}$ \ergs, the group of outliers seems to be consistent with the track followed by \1h\  with a steeper correlation
index of $\sim$ 1.4 \citep{Coriat11}, and also quite similar to the few reported neutron star systems \citep{Migliari06b}. In any-case, this larger dataset confirms the existence of two
tracks in the radio/X-ray flux correlation at higher X-ray flux, but still in the hard state (recently statically proved by \citealt{Gallo12}). 
It is unclear for the group of outliers if the normalisation of the correlation is similar  or not, as some dispersion is also observed along the \1h\ track. 
Furthermore, as found by  \citet{Coriat11, Coriat11b} and reinforced here  with the increased  number of sources  (e.g. \maxisz\ and \xteds\ \citealt{Ratti12, Jonker12}, Brocksopp et al. to be submitted), 
at luminosity below 4 $\times$ 10$^{36}$ \ergs\ the outliers seem to leave  the \1h\ track and join the  standard  correlation around 10$^{35}$ \ergs .

The spin of the BH \citep{Fender10}, as well as the binary  parameters (orbital period, accretion disc size, inclination, \citealt{Soleri11}) do not seem to
play a role in defining the level of jet power. By using a jet toy-model with an  increasing bulk Lorentz factor above 10$^{-3}$ \ledd, \citet{Soleri11} were able 
to reproduce the larger scatter for the brighter hard state (the zone of the outliers).  It is, however, clear from  Figure~\ref{fig_correlAll}  that 
the outliers are  a separate population from the sources lying on the standard correlation (see also \citealt{Gallo12}). The radio/X-ray diagram is not uniformly populated, as would
have been expected if the two populations were related to a uniform distribution of inclination angle (and henceforth Lorentz factor). The existence of a critical magnetic field \citep{Casella09}  
has also been invoked to explain the lower radio luminosity for the outliers, but then it also needs to explain the existence of  two populations as well as the steeper
correlation index. In order to find an explanation of the outliers by a  lower radio luminosity (or jet power), it may then be necessary to invoke a different 
coupling between jet power and mass accretion rate (see \citealt{Coriat11} for more details). 

Instead of characterising the outliers by fainter jet power (as in \citealt{Soleri11}), one may instead consider these sources as being more X-ray bright than the
ones from the  standard correlation, meaning that the understanding  of the correlations would be related to the origin of the X-ray emission process(es).  We would  
then be left with two groups of sources above 4 $\times$ 10$^{36}$ \ergs\ (one with an index of $\sim$ 1.4 and the other one with the typical index of  $\sim$ 0.6). 
It seems apparent that only the $\sim$ 0.6 track exists at lower radio and  X-ray luminosities. \citet[][and references therein]{Coriat11} have extensively discussed 
the nature of these two groups.  In that framework, the outliers would be consistent with the presence of radiatively efficient BHs in the hard state
(hot accretion flows or accretion disc corona). The standard correlation would still be related to a radiatively inefficient accretion flow (such as X-ray jets or 
ADAF  models).   The difference between the two populations could  possibly be related to variation in the $\alpha$ viscosity parameter \citep{Xie12}. \\

\subsubsection{Caveats on constraining the slope of the correlation using portion of one individual outburst }

The global results presented in this study for  \gx\  or in \citet{Coriat11}  for  \1h\  allow us to characterise the evolution of the radio and X-ray fluxes correlation 
for a given source along very different outbursts. While the global correlation is maintained 
with a specific slope ($\sim$  0.6  for \gx\  or $\sim$ 1.4 for  \1h), we note that {\it temporary deviations  do exist}. This happens for \gx\ several times, mostly in the decaying hard 
state once the compact jets have been fully rebuilt (see 2005, 2007 and 2011 for the decay), but  also  possibly even at higher fluxes (although it is not so clear at present,
see end of rises in 2002, 2007 and 2010). During these limited portions of outburst, correlation indices in the 0 to 2 range can  temporarily be observed.
It also happens to  \1h\  which shows two tracks and a transition (with a flatter index, \citealt{Jonker10,Coriat11}) between the two tracks. 
Additionally, we note that other sources have also been observed with different correlation indices (e.g. \gro\ \citealt{Shaposhnikov07, Xue07}). This concerns
Galactic BHs (and probably also accreting neutron star binaries). 

The observed variations of the correlation index in \gx\ (see Fig. \ref{fig_correl_separate}) occur over a 3--9 keV flux range of roughly a decade.  Several reasons could be invoked 
to explain temporary variations of the correlation index, such as changes in the radio spectral index or the radio to near infrared spectral index (as discussed in section 4.1), an increase of the importance of electrons  cooling in the jets, the presence of an additional X-ray emitting component, etc. 
In order to properly constrain the global correlation index, the 3--9 keV (or a similar X-ray band) flux should be sampled over more than a decade in X-ray flux, and probably observations over two decades 
in X-ray flux should provide a reasonable estimate of this index (e.g. see the example of \1h, \citealt{Coriat11}). However, if the source is an outlier (section \ref{sect_outliers}), special 
care should  be taken (with regular monitoring) in order to separate the different tracks (mixing the 1.4 track with observations taken during the transition to the standard 0.6 track could lead 
to  a different index).

As the correlation has been extended to AGNs \citep{Merloni03, Falcke04}, it 
also means that very different correlation indices could potentially  be measured in supermassive BHs with very  limited datasets (e.g. \citealt{King11}). 
Therefore, one has to be very  careful before  generalising  the slope of the correlation from a very specific portion of an outburst. Only very long studies
sampling  large radio and X-ray flux variations of a given source (or a large sample of different sources with known distance) could reveal  the intrinsic empirical relations 
necessary  to understand the coupling between accretion and ejection mechanisms in accreting BHs.

\subsection{On the nature of  ``hysteresis" in OIR }

During the 2000 outburst of \xte,   \citet{Russell07b}  plotted  the infrared versus X-ray emission and
reported quasi-parallel tracks for the rising and decaying hard states.  This phenomenon, that they 
called ``hysteresis'', has  been reported to date only in \xte. Whereas, the difference in normalisation 
between the rise and decay is of a factor 5 in \xte\  \citep{Russell07b}, \citet{Coriat09} already pointed out that {\it no 
difference} could be seen in infrared between the rise and decay of  \gx, despite an intense 
monitoring over 5 years. However, as discussed in Section \ref{sect_indiv}, we found some differences in radio
between the decaying and rising hard states. This occurs roughly during period of formation and destruction 
of the compact jet (similarly to \xte ). Although, we clearly do not have parallel tracks in the case of \gx , we  
possibly observed changes up to a factor 2 in term of variation of the normalisation of the radio track. Similarly
to \xte, the jet radio emission in \gx\ is sometimes  brighter during the decay compared to the rise for a given 
X-ray luminosity.  The radio observations conducted during the 2000 outburst of \xte\ \citep{Corbel01} are too 
sparse to be compared with the infrared data.  
 
To explain the ``hysteresis'' in \xte, \citet{Russell07b} considered changes in the radiative efficiency of 
the accretion flow, variations in the viscosity parameter $\alpha$ or  modification in the jets properties. 
Amongst these different scenarios, one needs to find one that could explain the limited ``hysteresis'' in \gx\ (only a 
factor 2 in radio and no difference in near-infrared). In \gx\ and \xte , the radio and infrared emissions in  the hard states 
originate from the compact  jets. The fact that no difference is observed in the near-infrared for \gx\ possibly implies that the origin
of this ``hysteresis''  is not solely related to the X-ray emission process(es), but rather to the compact jets. 

As discussed in Section \ref{sect_fdlir}, the radio, near-infrared and X-ray emissions  of \gx\ are strongly coupled.
It is then possible that this ``hysteresis'' is only determined by the spectral energy distribution of the compact jets and 
therefore traces the jet behaviour (rather  than the accretion flow). The level of ``hysteresis''  and the involved 
spectral domain could possibly be related to the  evolution of the break frequency where the jets becomes optically thin, 
but also to  the evolution of radiative efficiency  along the jets, the interplay between the jets and the corona and/or  the  
efficiency of particles acceleration. Alternatively the magnetization of the disk \citep{Petrucci08} could possibly modify 
the broadband spectral energy distribution of the jets.

\section{Conclusions}

We reported  a series of radio and X-ray observations of the recurrent BH  \gx\ during the past 15 years. This new set of observations 
samples almost all luminosity levels (nearly 5 decades in X-ray flux), possibly down to quiescence,  of a BH  along several different outbursts. \gx\ can therefore be considered
a reference  for comparison  with other accreting sources without any bias from distance uncertainties. 

Despite numerous outbursts of different nature, the overall  radio and X-ray emissions  display a very strong non-linear 
correlation with a coupling of the form L$_X \propto L_\mathrm{Rad}^{0.62 \pm 0.01}$. This very well constrained correlation index is consistent with previous studies, that 
highlighted an index in the range of 0.5 to 0.7. The same correlation seems to be maintained from the brightest hard states to the lowest detected luminosity level,
that could be consistent with the quiescence state of \gx.   With the reported long-term near-infrared/X-ray fluxes correlation in \gx\ \citep{Coriat09},
we further demonstrated a coupled correlation between these three frequency ranges. The level of radio emission could then be tied to the near-infrared 
emission by the evolution of the jet broad band SED (assuming that the near-infrared data lies on the optically thick portion of the compact jet spectrum).

We also highlighted periods of  significant  and higher dispersion along the fitted function in the radio/X-ray   sample, when compared to the near-infrared/X-ray data.
This appeared more pronounced in periods of formation and destruction of the self absorbed compact jets. For a given X-ray flux, the radio emission could be higher 
(up to a factor 2) during the decay compare to the rise of the outburst, whereas no difference was found in near-infrared \citep{Coriat09}. By comparing this behaviour
with a similar one in \xte\ \citep{Russell07b}, this ``hysteresis'' behaviour seems to  be related to the properties of the compact jets.

We incorporated our new data of \gx\ in a more global study that includes a set of 24 BHCs. We observed that  the ``standard'' track, that is now well constrained  by \gx , is 
closely shared by a subset of these BHCs (e.g. \vq ). The quiescent BH  \a0\   is the faintest reported source (more than 2 orders of magnitude fainter than
\gx  , \citealt{Gallo06}) and it obeys the standard track that is defined with our large \gx\  sample. Furthermore, we collected a growing number of sources (the majority in fact) 
that  appear significantly below the ``standard'' track (see also \citealt{Gallo12}), those ``outliers'' may possibly share the behavior reported for \1h, meaning that they could join the 
standard correlation at low and/or high X-ray fluxes. We emphasize  that special care should be taken to constrain the index of the correlation in 
very limited sample, as this index  can only be securely  obtained with more global  studies such as the one we conducted with \gx.  
The location of a  source in a given track of the radio/X-ray diagram could  be related to the radiative efficiency 
of the inner accretion flow or a different coupling between jet power and mass accretion rate \citep{Coriat11}. While the standard track has been extended to supermassive BHs
by the so-called ``fundamental plane of BHs activity''  \citep{Merloni03,Falcke04}, the existence a population similar to the ``outliers'' within AGNs would be a strong 
support  for scale invariance of the jet-accretion coupling in accreting black holes \citep[see recent work of][]{Broderick11}.

\section*{Acknowledgments}

The authors would like to thank the anonymous referee for the careful reading of the manuscript. 
The Australia Telescope is funded by the Commonwealth of Australia for operation as a national
Facility managed by CSIRO. The research leading to these results has received funding from the 
European CommunityÕs Seventh Framework Programme (FP7/2007-2013) under grant agreement number ITN 215212 ÓBlack Hole Universe.
JAT acknowledges partial support from   NASA {\em Swift} Guest Observer grant NNX10AK36G and also from
the NASA Astrophysics Data Analysis Program grant NNX11AF84G. MMB and CDB are supported by NSF/AST grants 0407063 and 070707 to CDB. 
We acknowledge the use of data obtained from the High Energy Astrophysics Science Archive Research Center (HEASARC), provided by NASA's Goddard
Space Flight Center. We thank  the ISSI in Bern where an earlier version of this work was discussed. 
SC would like to thank Sera Markoff, James Miller-Jones, Dave Russell, Wenfei Yu, Feng Yuan, Andrzej Zdziarski  for discussions, J\'erome Rodriguez for a careful reading of 
the manuscript, Jeroen Homan and Tomaso Belloni for sharing informations on the
X-ray states of \gx\ during the past years, Craig Markwardt  for informations on the RXTE Galactic Bulge monitoring and Anthony Rushton for sharing the data of \grs.

\bibliographystyle{mn2e_fixed}


\end{document}